\documentclass[reprint, superscriptaddress, amsmath, field, amssymb, aps, prl]{revtex4-2}
\usepackage{xcolor}
\usepackage{lipsum} 
\usepackage[version=4]{mhchem}
\usepackage{appendix}
\usepackage{braket}
\usepackage{tikz}
\usepackage{booktabs}
\usepackage{amsmath}
\usepackage{comment}
\usepackage{amssymb}
\usepackage{mathtools}
\newcommand{\dd}{\mathrm{d}}
\usepackage{chemformula}
\usepackage{graphicx}
\usepackage{dcolumn}
\usepackage{bm}
\usepackage{setspace}
\usepackage{physics}
\usepackage{braket}
\usepackage[colorlinks=true, citecolor=blue, linkcolor=blue, urlcolor=blue]{hyperref}
\usepackage{indentfirst}
\usepackage{cleveref}
\usepackage{bbm}
\usepackage[english]{babel}

\usepackage[thinc]{esdiff}

\definecolor{rungred}{HTML}{C25450}
\definecolor{ratiogreen}{HTML}{00917B}
\definecolor{skewyellow}{HTML}{AB9D05}

\newcommand{\markerred}{\tikz[baseline=-0.6ex]\draw[rungred,fill=rungred] (0,0) circle (3.5pt);}
\newcommand{\markergreen}{\tikz[baseline=-0.6ex]\draw[ratiogreen,fill=ratiogreen,rotate=45] (-2.5pt,-2.5pt) rectangle (2.5pt,2.5pt);}
\newcommand{\markeryellow}{\tikz[baseline=-0.6ex]\draw[skewyellow,fill=skewyellow] (-3pt,-3pt) rectangle (3pt,3pt);}

\definecolor{rungpink}{RGB}{126,100,213}
\newcommand{\markerpink}{\tikz[baseline=-0.5ex]\draw[rungpink, fill=rungpink, dashed] (-3.5pt,-2.5pt) -- (0,3.5pt) -- (3.5pt,-2.5pt) -- cycle;}

\def \ETH{Institute for Theoretical Physics, ETH Z\"urich, CH-8093 Z\"urich, Switzerland}

\newcommand{\pheader}[1]{{\emph{#1}}\textemdash}

\newcommand{\avg}[1]{\left\langle #1 \right\rangle_{0}}

\def \ETH{Institute for Theoretical Physics, ETH Z\"urich, CH-8093 Z\"urich, Switzerland}
\def \MaxPlanck{Max-Planck-Institut f\"ur Quantenoptik, 85748 Garching, Germany}
\def \Munich{Munich Center for Quantum Science and Technology (MCQST), 80799 Munich, Germany} 
\def \LMU{Fakult\"at f\"ur Physik, Ludwig-Maximilians-Universit\"at, 80799 Munich, Germany}
\def \IBM{IBM Quantum, IBM Research -- Zurich, 8803 R\"uschlikon, Switzerland}

\begin{document}
\title{Characterizing quasiparticles in strongly correlated systems \\using nonlinear spectroscopy in quantum simulators}

\author{Luka Skolc}
\thanks{These authors contributed equally to this work}
\affiliation{\ETH}

\author{Utso Bhattacharya}
\thanks{These authors contributed equally to this work}
\affiliation{\ETH}
\affiliation{\IBM}

\author{Jonathan B. Curtis}
\affiliation{\ETH}

\author{Immanuel Bloch}
\affiliation{\MaxPlanck}
\affiliation{\Munich}
\affiliation{\LMU}

\author{Eugene Demler}
\affiliation{\ETH}

\date{\today}
\begin{abstract}
Characterizing carrier type in strongly correlated quantum systems conventionally relies on the Hall effect.
In cold-atom quantum simulators, implementing Hall transport measurements typically requires synthetic gauge fields, which introduces significant heating.
Here, we present nonlinear spectroscopic and fluctuation-based protocols that establish an alternative route to identifying the sign of quasiparticle charge. 
We demonstrate that second-order density and current responses to finite-momentum quenches and drives—as well as equilibrium third-order density cumulants—distinguish electron- from hole-like quasiparticles.
For a Fermi–Hubbard multi-leg ladder on a square lattice, numerical simulations reveal a crossover from hole- to electron-like carriers upon hole-doping away from half-filling, matching the sign change in the Hall coefficient.
To demonstrate the applicability of our protocols beyond fermionic systems, we show that the nonlinear density response to a finite-momentum, finite-frequency drive reveals the sign of charge carriers in a hard-core boson ladder, and that the nonlinear signal can be significantly enhanced when driving near resonance with a nonlinear collective mode.
Our results establish nonlinear density and current response and equilibrium non-Gaussian fluctuations as complementary probes, offering quantum simulators a direct route to characterize the carrier sign, without the experimental hurdles of conventional transport setups.
\end{abstract}
\maketitle

\pheader{Introduction}
Measuring the Hall resistance is a fundamental experimental technique for determining charge carrier type across semiconductors, metals, liquids and plasmas. One of the earliest puzzles in solid-state physics was the varying sign of the Hall coefficient, $R_H$. In the semiclassical model of transport, a system of particles with charge $q$ and number density $n$ yields $R_H = 1/nq$, which explicitly identifies the sign of the charge carriers. This contrasts with the Drude formula for longitudinal conductivity, which depends on $q^2$ and is insensitive to the carrier sign. Because early experiments suggested that current in solids is carried by negatively charged electrons, subsequent measurements of positive Hall coefficients in certain metals came as a surprise.
The paradox was resolved by the quantum-mechanical description of electrons in periodic potentials, which led to the band theory framework and introduced the concept of holes as charge carriers.

Ensuing studies of strongly correlated electron systems have revealed new puzzles in Hall physics. In high-$T_c$ cuprates, for instance, $R_H$ undergoes multiple changes in both sign and magnitude across different doping regimes~\cite{badoux_change_2016, PhysRevB.83.054506, leboeuf_electron_2007}, in direct contradiction with band structure predictions. However, there is currently no unified theoretical understanding of the Hall effect in strongly correlated electron systems. In cuprates, interpretations for the sign and magnitude changes in $R_H$ range from Fermi surface reconstruction due to charge ~\cite{badoux_change_2016, PhysRevB.83.054506, leboeuf_electron_2007}, spin~\cite{harrison_spin-density_2009}, or hidden $d$-wave density-order~\cite{chakravarty_fermi_2008}, to emergence of new types of quasiparticles in doped Mott insulators (see calculations using the DMFT approach \cite{kuchinskii_hall_2022} and phenomenological Yang-Rice-Zhang ansatz \cite{PhysRevB.73.174501, storey_hall_2016}), and to scenarios based on fractionalized topological Fermi liquids \cite{sachdev_novel_2016}. 
Hall anomalies have also been observed in heavy-fermion~\cite{Paschen.2004, Hundley.2004}, transition-metal dichalcogenide~\cite{Evtushinsky.2008}, and moir{\'e} materials~\cite{Ghiotto.2024}, suggesting that these phenomena are a common hallmark of strong electronic correlations.

These surprises in Hall physics of correlated materials have motivated studies within simplified model Hamiltonians.
In particular, the paradigmatic Fermi Hubbard model that is widely regarded as the minimal framework for capturing Mott physics and its descendant correlated phases \cite{arovas_hubbard_2022, lee_doping_2006} has been a focal point of theoretical investigations.
While analytic approaches and state-of-the-art numerical calculations\textemdash such as thermodynamic expansions \cite{khait_hall_2023} and determinant quantum Monte Carlo \cite{wang_dc_2020, wang_numerical_2021}\textemdash have revealed a sign reversal at intermediate dopings in the 2D Fermi-Hubbard model for temperatures comparable to the antiferromagnetic temperature scale $J$ and higher, the notorious fermion sign problem and convergence limitations prevent these methods from reaching the low temperatures relevant to cuprate experiments.

\begin{figure}[t]
        \centering
        \includegraphics[width = 0.99 \linewidth]{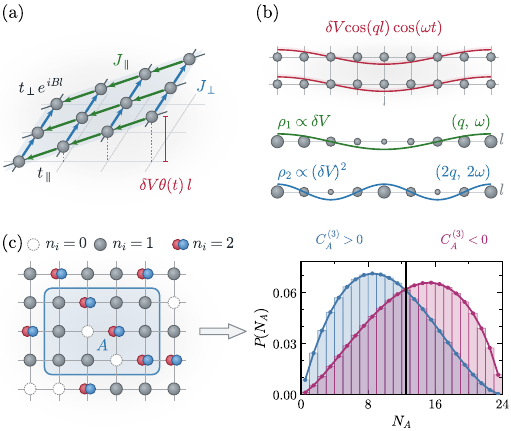}
        \caption{Illustration of the measurement protocols.
        (a) Hall response: a potential gradient $\delta V$ is quenched across a 2D lattice (hopping amplitudes $t_\parallel$, $t_\perp$) in a magnetic field $B$. The quench generates longitudinal currents $J_\parallel$, and the induced transverse currents $J_\perp$ determine the Hall coefficient $R_H$, which reveals the quasiparticles' charge \cite{PhysRevLett.126.030501}.
        (b) Second-order density response to a quench or time-periodic drive of a spatially modulated onsite potential. Periodic driving with frequency $\omega$ and momentum $\mathbf{q}$ induces first-order density response $\rho_1$ at $(\omega, \mathbf{q})$ and  nonlinear response $\rho_2$ at $(2\omega,2\mathbf{q})$. The sign of $\rho_2$ determines the sign of charge carriers.
        (c) Skewness $C_A^{(3)}$ of equilibrium density fluctuations in a subsystem $A$ can be found from repeated projective measurements. Positive-skewed distributions (blue histogram) imply electron-like quasiparticles; negative-skewed (red) imply hole-like. The vertical black line is the common mean of the distributions.}
        \label{fig:protocols} 
\end{figure}

New advances in fermionic cold-atom quantum simulators now offer the optimal testbed for exploring the low-temperature physics of the Fermi-Hubbard model~\cite{Xu.2025}, and have already provided significant insight into the magnetic fluctuations and the interplay of spin and charge degrees of freedom in the pseudogap regime~\cite{Kendrick.2025,Chalopin.2026,Andrei.2026}. 
To directly simulate the Hall effect with neutral atoms, one needs to introduce a synthetic magnetic gauge field.
Floquet engineering has enabled studies of lattice models with strong magnetic fluxes \cite{aidelsburger_experimental_2011, atala_observation_2014,aidelsburger_realization_2013, PhysRevLett.111.185302,jotzu_experimental_2014}, and was recently extended to the strong-interaction regime in systems of bosons~\cite{impertro_strongly_2025} and fermions~\cite{leonard_realization_2023}.
Earlier theoretical works have proposed measuring the Hall effect in cold atom systems by quenching a potential gradient (see Fig. \ref{fig:protocols}(a) for a schematic) \cite{PhysRevLett.126.030501}.
However, the periodic driving required to generate the synthetic magnetic fields inherently induces heating and decoherence in the atomic system, consequently restricting state lifetimes, complicating the state preparation in the presence of a synthetic field, and inhibiting steady-state transport measurements like the ones done in the absence of synthetic fields \cite{brown_bad_2019, guardado-sanchez_subdiffusion_2020}. We remark that while recent experiments have measured the Hall effect in synthetic atomic ladders \cite{zhou_measuring_2025}, extending this internal-state approach to mimic fully two-dimensional optical lattices remains challenging.

In this work, we introduce a series of novel protocols for measuring the charge of quasiparticles in cold-atom quantum simulators, demonstrating that carrier sign and Hall-like anomalies can be resolved without steady-state transport in a synthetic gauge field.
Our protocols utilize the experimentally accessible single-site resolution of density and current \cite{sherson_single-atom-resolved_2010,PhysRevLett.133.063401} to extract higher-order response and correlation functions \cite{hofferberth_probing_2008, sanner_suppression_2010,Muller.2010, sanner_speckle_2011,Lebrat.2024,Chalopin.2026exl}, which\textemdash as we demonstrate\textemdash are highly sensitive probes of quasiparticle character.
With readily tunable system parameters and full counting statistics at hand \cite{Wang.2024ete,PhysRevA.75.063611,schweigler_experimental_2017,Chalopin.2026exl}, cold-atom platforms are perfectly posed to characterize emergent quasiparticles in the Fermi-Hubbard and related models and resolve puzzles about the Hall anomalies.

\pheader{Protocols}
The Hall effect can distinguish carrier types because it involves a nonlinear, even-power response function connecting the induced transverse field $E_y = R_H J_x B_z$ to the product of the applied current density and magnetic field.
This motivates us to explore two categories of nonlinear measurements\textemdash second harmonic generation (SHG) response functions and density skewness correlations\textemdash as illustrated in Fig.~\ref{fig:protocols}(b,c).

We first consider the second-order density response, which in a conventional solid is defined in terms of the induced polarization {\it via} $\chi^{(2)}_{jkl}(2\omega; \omega,\omega) = \frac{\delta^2 {P}_{j}(2\omega ) }{\delta {E}_k(\omega)\delta E_l(\omega) } \propto e^3 $.
Since $\chi^{(2)}$ involves an odd power of the quasiparticle charge, as opposed to e.g. resistivity, it can in principle distinguish between electron- and hole-like response. 
While this response is strictly forbidden at zero total momentum ($\mathbf{q}=0$) in centrosymmetric systems by inversion symmetry, it becomes allowed at finite momenta $q = (\omega,\mathbf{q})$, where ${P}_{j}(2q) = \chi^{(2)}_{jkl}(2q; q,q){E}_k(q)E_l(q)$~\cite{sun_universal_2018}. However, the finite-$\mathbf{q}$ effect is practically unobservable in the far field optical experiments on solids as the photon momentum is negligibly small compared to typical momenta in the solid.
Crucially, this is not the case in cold-atom quantum simulators, which can be perturbed and probed with single-site resolution.
As we will demonstrate, the carrier type of quasiparticles can be resolved by measuring the second-order density response to either a quench or a periodic drive of a spatially modulated onsite potential.
We remark that recent experiments with near-field probes have also begun probing finite-momentum nonlinear conductivities of solids \cite{nearfield2021}.

The second class of protocols we propose utilizes the unique access to higher-order counting statistics afforded by the single-site resolution available in cold-atom quantum simulators~\cite{Bakr.2010}. 
We will show that measuring the \emph{equilibrium fluctuations} of the number of particles in a subsystem can be used to probe the carrier type. Specifically, we will study the grand-canonical skewness~\cite{Armijo.2010} (third cumulant) $C^{(3)} \equiv\frac{1}{V}\sum_{jkl\in V }C^{(3)}_{jkl}$ of the number of particles in a subsystem of volume $V$, where the connected correlation function $C^{(3)}_{jkl} = \langle \hat{n}_j\hat{n}_k\hat{n}_l\rangle_c$ and $\hat{n}_j = \sum_\sigma \hat{a}_{j\sigma}^\dagger \hat{a}_{j\sigma}$ is the density of particles on site $\mathbf{r}_j$.
In analogy with the linear fluctuation-dissipation theorem which connects first-order response functions to quadratic equilibrium fluctuations, nonlinear response functions such as the second-harmonic density response are related to higher-order correlation functions~\cite{Wang.2002}.
This motivates studying non-Gaussian density fluctuations as a complementary probe to $\chi^{(2)}$ \cite{Curtis.2025}.
A key advantage of this protocol is that it entails only equilibrium measurements, avoiding the heating due to quenching or periodically modulating the system.

Our protocols are generically applicable to fermionic or bosonic cold-atom systems. In this paper, we first illustrate their utility on the Fermi-Hubbard model. We compare the results to the Hall response and show the various approaches are in agreement and faithfully resolve the changes in quasiparticle physics. We then switch to bosonic systems to explore second harmonic-order density response in greater detail and provide analytic results.

\pheader{2D Fermi-Hubbard model}
The Fermi-Hubbard model (FHM) is a paradigmatic model for strongly-correlated systems such as high-$T_c$ cuprate superconductors~\cite{Zhang.1988,Emery.1987}, as well as cold atoms in optical lattices~\cite{Mazurenko.2017}. It is described by the Hamiltonian 
\begin{equation}
    \label{eqn:FHM}
    \hat{H} = -t\sum_{\langle j,k\rangle}\sum_{\sigma} \left(\hat{c}^\dagger_{j\sigma}\hat{c}_{k\sigma} + \textrm{h.c.}\right) + U \sum_{j} \hat{n}_{j\uparrow} \hat{n}_{j\downarrow},
\end{equation}
where $t$ is the nearest-neighbor hopping amplitude between adjacent sites $\langle j,k\rangle$ on a square lattice with $V = L\times M$ sites in total, where there are $M$ legs with $L$ sites each, $U$ is the onsite Coulomb repulsion, $\hat{c}^\dag_{j\sigma}$ creates a fermion with spin $\sigma \in {\uparrow,\downarrow}$ at lattice site $j$, and $\hat{n}_{j\sigma} = \hat{c}_{j\sigma}^\dagger \hat{c}_{j\sigma}$ is the density of spin $\sigma$ on site $j$.

Despite its simple form, the equilibrium phase diagram of FHM as a function of temperature $T$, Coulomb interaction $U$, and hole-doping fraction $p = 1 -N/V$ (expressed in terms of total number of fermions $N$ and total system volume $V$), is extremely complicated and subject of much debate~\cite{Andrei.2026,Qin.2019,Wietek.2021}. 
We focus on the low-temperature strongly-correlated regime with $T \ll t \ll U$ where different theoretical approaches~\cite{khait_hall_2023} have suggested the Hall coefficient should change sign with increasing $p$.

\begin{figure}[t]
        \centering
        \includegraphics[width = 0.95\linewidth]{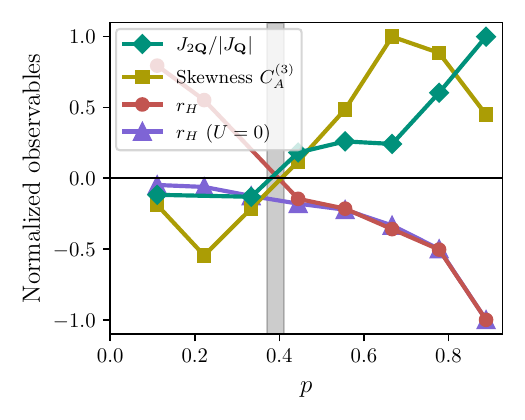}
    \caption{
    Data obtained using TDVP on a $6\times3$ Fermi-Hubbard ladder with $U = 10t$. Three independent protocols reveal a quasiparticle transition at critical hole doping $p_c \approx 0.37 - 0.41$ (gray shaded region), signaling a change from hole-like ($p < p_c$) to electron-like ($p > p_c$) carriers.  [\protect\markerred] Dimensionless Hall response coefficient $r_H$ (see SM).  [\protect\markergreen] Normalized second-order current response $J_{2\mathbf{Q}}/|J_\mathbf{Q}|$ following a modulated potential quench ($V = 0.45t$) with $\mathbf{Q} = 2\pi/L\hat{e}_x$. [\protect\markeryellow] Equilibrium density skewness $C_A^{(3)}$ measured in a central subsystem. 
    [\protect\markerpink] Hall response for non-interacting fermions ($U = 0$). 
    All observables are normalized to facilitate comparison across protocols. $B$ and $ \mathbf{Q}$ are measured in lattice units.}
    \label{fig:fermi_hubbard} 
\end{figure}

To explore this sign reversal, we study both the equilibrium properties and quench dynamics of a $6\times 3$ spin-balanced $(N_\uparrow=N_\downarrow)$ Fermi-Hubbard multi-leg ladder with $U/t=10$. The ground state is obtained using the density matrix renormalization group (DMRG), while the subsequent real-time dynamics are simulated using the time-dependent variational principle (TDVP). Both calculations employ matrix-product state representations with a maximum bond dimension of $\chi_{\rm max}=10,000$ (see the SI for further details). The results are shown in Fig.~\ref{fig:fermi_hubbard}.

To benchmark our protocols, we first simulate the equilibrium dimensionless DC Hall coefficient $r_H\equiv e_0 R_H / V$ by applying a persistent current along the lattice legs (long dimension), in the presence of a static magnetic field threading the lattice (see SM). 
Strikingly, $r_H$ reverses direction around a critical doping $p_c \approx 0.4$ in the strongly-interacting regime, switching from hole-like behavior at low dopings $p<p_c$ to an electron-like response at $p > p_c$. In contrast, $r_H$ always remains electron-like for $U=0$.

Alternatively, the sign of charge carriers can be probed by either studying the early-time, second harmonic (in momentum) of the density response after a quench, or the steady-state second-harmonic response (in momentum and frequency) under a continuous-wave drive.
We give an argument why these two responses are closely related measures of quasiparticle charge in Appendix B.
Inspired by quench-type transport experiments in the FHM \cite{brown_bad_2019}, we focus here on rapidly switching on a spatially modulated potential 
\begin{equation}
    \hat{H}_{\rm pot}(t) = \Theta(t)\sum_{j}V\cos(\mathbf{Q}\cdot \mathbf{R}_j) \hat{n}_j,\label{eq:potential}
\end{equation}
which induces out-of-equilibrium dynamics.
$\mathbf{R}_j$ denotes lattice site positions and $\Theta(t)$ is the Heaviside function. 
We track the evolution of the longitudinal current 
\begin{equation}
   J_{\bf q}(t) = \sum_{\langle j,k \rangle} e^{-i\mathbf{q}\cdot\mathbf{R}_{j,k}} \langle \hat{J}_{j,k}(t) \rangle, 
\end{equation}
where $\hat{J}_{j,k} = it\sum_\sigma(\hat{c}_{j\sigma}^\dagger \hat{c}_{k\sigma} - \text{h.c.})$ is the bond current operator and $\mathbf{R}_{j,k}$ connects neighboring sites.
While the current response is broadband in frequency, its momentum-space structure can be resolved using the single-site resolution available in quantum gas microscopes.
The quench in Eq. \eqref{eq:potential} induces both linear ($\mathbf{q}= \mathbf{Q}$) and nonlinear ($\mathbf{q}= 2\mathbf{Q}$) response in both the local density and current.
We quantify the nonlinearity by the ratio $J_{2\mathbf{Q}}/|J_\mathbf{Q}|$ of the second harmonic to the norm of the fundamental current response at an early time $t_0$ (details in SI). Whereas the sign of $J_\mathbf{Q}$ is insensitive to filling, the second-order response reverses sign at $p\approx p_c$, just like the Hall coefficient.

Finally, we observe that skewness of the density fluctuations  within a small subsystem  also changes sign at the critical doping $p_c$. For a subsystem $A$, the skewness is defined as  
\begin{equation}
C^{(3)}_A = \frac{1}{V_A} \langle (\hat{N}_A - \langle \hat{N}_A \rangle)^3 \rangle,
\label{eqn:skew}
\end{equation}
where $\hat{N}_A = \sum_{j \in A} \hat{n}_j$ is the number of particles in the subsystem $A$ and $V_A$ is the number of sites.
Here, $A$ is chosen as a central, contiguous block that excludes boundary sites, thereby isolating bulk behavior (see SI for details). 
Physically, this quantity probes the asymmetry of the particle-number Fock state distribution, as expanded on later in the finite-temperature section. 
Importantly, such subsystem-resolved density statistics are directly accessible in cold-atom experiments~\cite{Armijo.2010}.
Having established the numerical validity of our protocols for detecting changes in the sign of quasiparticles in the FHM, we now analyze them in greater detail, starting with second-order density response.

\begin{figure}[t]
        \centering
        \includegraphics[width = 0.99 \linewidth]{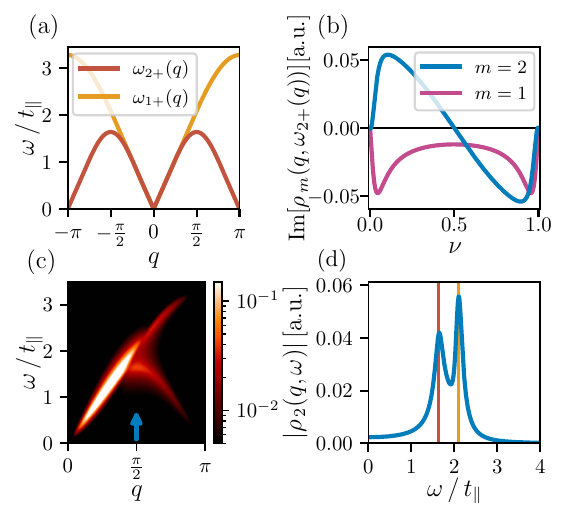}
        \caption{
        SHG in a HCB ladder. (a) Dispersion of linear (orange) and nonlinear (red) collective modes $\omega_{1+}(q)$ and $\omega_{2+}(q)$. Data shown for transverse hopping $t_\perp=0.25t_\parallel$, driving strength $\delta V =0.2t_\parallel$, damping rate $\eta=0.1t_\parallel$ and filling $\nu=0.25$. (b) Imaginary part of first- (magenta) and second-order (blue) density response functions, for a driving wavevector $q=\pi/2$, evaluated at the nonlinear resonance $\omega =\omega_{2+}(\pi/2)$ as a function of filling $\nu$. At the half-filling point, $\rho_2(q,\omega)$ changes sign, whereas $\rho_1(q,\omega)$ is invariant under the particle-hole transformation. (c) Magnitude of second-order density response $\vert \rho_2(q,\omega)\vert$ as a function of drive frequency and wavevector. $\rho_2(q,\omega)$ is enhanced when the drive matches  resonance curves from (a). By driving at large $q$, we can selectively address the nonlinear resonance while keeping $\rho_1(q,\omega)$ small. The blue arrow denotes $q=\pi/2$. (d) Nonlinear density response $\vert \rho_2(q,\omega)\vert$ at fixed wavevector $q=\pi/2$ is enhanced when $\omega =\omega_{1+}(\pi/2)$ or $\omega =\omega_{2+}(\pi/2)$. The y-label is shared between plots (c) and (d). Anisotropic hopping was found to boost the nonlinear peak in second-order response; see SI for data with $t_\perp=t_\parallel$. }
        \label{fig:shg_main} 
\end{figure}

\pheader{Second harmonic generation}
The periodic potential quench discussed in the \textit{Protocols} section probes response at fixed wavevector but offers no frequency resolution. 
In this section, we present a related \textit{Second Harmonic Generation} (SHG) protocol where the atoms are driven by a space- and time-periodic potential modulation with a well-defined wavevector $\mathbf{q} $ and frequency $\omega$. 
In the steady state, current and density modulations will occur at $(\mathbf{q},\omega)$ to first order and at $(2\mathbf{q},2\omega)$ to second order of the drive strength. 
We will see that driving at a single frequency allows us to selectively address collective modes of first- or second-order fluctuations, thereby boosting the relative strength of the nonlinear response and improving the signal-to-noise ratio.  

To demonstrate the broad applicability of our protocols outside of fermionic  and to establish analytical insight in a simpler model, we turn to bosonic platforms in this section.
Beyond cold-atom simulators where bosonic setups are generically easier to engineer and control, various bosonic Hamiltonians such as the Bose-Hubbard model, the XY model or the bosonic $t-J$ model can be realized in optical tweezers \cite{PhysRevLett.132.230401} or superconductng qubits \cite{karamlou_probing_2024,andersen_thermalization_2025}, which could also utilize our protocols.
Here, we demonstrate the SHG protocol on a two-leg square ladder of hard-core bosons (HCBs).
Since the Hall coefficient of HCBs in both ladder \cite{citro_hall_2025} and square-lattice \cite{lindner_conductivity_2010, PhysRevB.82.134510} geometries exhibits a sign reversal at half-filling, this system is an ideal platform to verify our SHG protocol correctly captures the transition between particle- and hole-like transport. 
In the absence of driving, 
\begin{equation}
    \hat{H} = -t_\parallel\sum\limits_{n,l}\left(\hat{b}_{n,l}^\dag \hat{b}_{n,l+1} + \mathrm{h.c.}\right) - t_\perp\sum\limits_l \left(\hat{b}_{1,l}^\dag \hat{b}_{2,l}+\mathrm{h.c.}\right),
\end{equation}
where $t_\parallel$ and $t_\perp$ represent the hopping amplitudes along the legs and rungs of the ladder, respectively. The operator $\hat{b}^\dag_{n,l}$ creates a boson on leg $n$ and rung $l$. The hard-core limit is enforced by the constraint $(\hat{b}_{n,l}^\dag)^2 = 0$, mapping the HCBs onto a spin-1/2 ladder. We introduce a time-dependent perturbation $\delta \hat{H} = \sum_{n,l} \delta V \hat{b}^\dag_{n,l} \hat{b}_{n,l} e^{i(ql - \omega t)} + \text{h.c.}$, representing a running-wave drive coupled to the local density. While standing-wave configurations are more readily implemented in experiment via dynamic superlattices, we utilize here a plane-wave drive to cleanly isolate the higher-harmonic response (see SI for the standing wave treatment).

In order to keep the calculation analytically tractable while capturing the essential physics, we employ a Gutzwiller variational wavefunction within the Dirac-Frenkel Lagrangian formalism \cite{10.21468/SciPostPhys.9.4.048}. We expand the non-equilibrium equations of motion around the ground state up to second order in the drive strength $\delta V$ (see SI for details). Since the steady-state modulations in density and current are connected by the Fourier-space continuity equation $J_\parallel(q,\omega) = \frac{i\omega}{1- e^{-iq}}  \rho(q,\omega)$, the density response serves as a complete proxy for the longitudinal transport dynamics. Within linear response
\begin{align}
    \rho_1(q,\omega)  =\frac{ \delta V t_\parallel 4\nu(1-\nu)  (1-\mathrm{cos}(q))}{\omega^2-\omega_{1+}^2(q) +2i\omega \eta},\label{eq:shg_rho1}
\end{align}
where the index $1$ denotes the perturbation order, is proportional to $\delta V$ and independent of the particle-hole nature of the filling $\nu$. The response is strongly enhanced when the drive frequency matches the dispersion $\omega_{1+}(q)$ of the symmetric collective mode corresponding to in-phase density oscillations along the legs. While the collective mode disperses linearly at long wavelengths as $\omega_{ 1+}(q\rightarrow 0) = |q|\sqrt{4 t_\parallel( 2t_\parallel +t_\perp) \nu(1-\nu)}$, it shows significant curvature as $q^{-1}$ approaches the lattice scale (see Fig. \ref{fig:shg_main}(a)). We shift all frequencies $\omega\rightarrow \omega+i\eta$ by an experiment-dependent damping coefficient $\eta$ that models trap inhomogeneity, thermal broadening and quasiparticle scattering. The second-order density response at $(2q,2\omega)$ (detailed in the SI)
\begin{equation}
\rho_2(q,\omega) \sim \frac{\delta V^2 t_\parallel^2}{ \omega^2 - \omega_{2+}^2(q) + 2i\omega \eta} \times \frac{\nu - 1/2}{ \omega^2 - \omega_{1+}^2(q) + 2i\omega \eta}\label{eq:shg_rho2}
\end{equation}
scales as $\delta V^2$ and, importantly, reverses sign under the particle-hole transformation. Equation \eqref{eq:shg_rho2} reveals that $\rho_2$ is resonantly enhanced not only at the fundamental mode $\omega=\omega_{1+}(q)$, but also at a nonlinear resonance $\omega_{2+}(q) \equiv \omega_{1+}(2q)/2$, characteristic of a two-wave mixing process. By tuning the drive to $\omega \approx \omega_{2+}(q)$, we can selectively boost the nonlinear signal while keeping the linear response $|\rho_1(q,\omega)|$ off-resonantly suppressed. We evaluate the imaginary part of $\rho_2$ at the nonlinear resonance for $q =\pi/2$ and compare it to $\rho_1$ across different fillings in Fig. \ref{fig:shg_main}(b). Not only are the signals comparable in magnitude, the second harmonic also exhibits a definitive sign change precisely at half-filling, demonstrating sensitivity to the particle-hole asymmetry of the HCB ladder.
This motivates SHG as a diagnostic for quasiparticle charge complementary to quench-type protocols.

Furthermore, as shown in Fig. \ref{fig:shg_main}(c), $\rho_2(q,\omega)$ develops distinct resonance peaks for $q \gtrsim \pi/2$, marking the regime where the $\omega_{1+}$ and $\omega_{2+}$ dispersion curves significantly diverge (see Fig. \ref{fig:shg_main}(a)). This underscores the necessity of driving at lattice-scale wavelengths while maintaining resonance with intrinsic collective modes, a task much easier for cold atoms than the solid state. Finally, for a drive wavevector $q=\pi/2$, Fig. \ref{fig:shg_main}(d) illustrates the substantial second-order response generated at momentum $\pi$. Obtaining this appreciable nonlinear signal requires strong driving, highlighting the distinct advantage of cold-atom quantum simulators, where $\delta V$ can be tuned to a significant fraction of the intrinsic energy scales. While the results shown in Fig. \ref{fig:shg_main} are for HCBs, coupling to nonlinear resonances at large drive momenta to resolve second-order response could be applied to fermionic systems as well.

\pheader{Skewness and finite-temperature effects}
Coming back to fermionic systems, we now analyze the density skewness in detail, with a particular emphasis on finite-temperature effects. 
To gain intuition, we first consider a noninteracting band theory in the grand-canonical ensemble, where the skewness of the particle number distribution is readily expressed in terms of the Fermi-Dirac occupation functions $f_{\bf k \alpha}$ (for details see SI).
We find 
\begin{equation}
    C^{(3)}= \int_{\bf k}\sum_{\alpha} f_{\bf k\alpha}(1-f_{\bf k\alpha})(1-2f_{\bf k\alpha}),
\end{equation}
which vanishes at full and empty filling, as well as at half-filling where it changes sign; for particle-like filling we expect $C^{(3)} >0$ and for hole-like $C^{(3)}<0$. 
Comparing to Fig.~\ref{fig:fermi_hubbard}, we see the exact results at $T = 0$ agree with band-theory in the overdoped regime of $p>p_c$ where the skewness is positive (particle-like). 
In contrast, exact results depart markedly from band theory in the underdoped regime of $p<p_c$, with negative skewness indicating hole-like carriers.
At finite temperature, it remains to be determined whether this sign change occurs at temperatures $T\sim U$ (as suggested by quantum Monte Carlo~\cite{Khait.2023}), which corresponds to the formation of magnetic moments, or at $T\sim J$ where magnetic correlations develop (as suggested by spin density-wave mean-field theory~\cite{Schrieffer.1989,Chi.1994,Peters.2014} -- see SI), or alternatively somewhere in between ($T \sim t$).
Nevertheless, the sign reversal in the skewness should be visible at temperatures which are accessible to current cold-atom setups~\cite{Chalopin.2026,Kendrick.2025,Xu.2025}, offering an opportunity for experiments to resolve this debate.

\pheader{Discussion}
In this work, we proposed a range of independent protocols to gain new insight into quasiparticle behavior in strongly-correlated systems by exploiting the local control and single-site resolution of cold-atom quantum simulators.
Specifically, we demonstrated how nonlinear probes\textemdash utilizing either quench or modulation spectroscopy\textemdash can extract the effective carrier charge in both the Fermi-Hubbard model and its bosonic analogs.
Similarly, we have shown that three-point cumulants~\cite{Lebrat.2024} can be used to probe the formation of hole-like magnetic polarons by capturing higher-order correlations within the Fock-space distribution.
The latter protocol can be employed \emph{in equilibrium}, making it particularly promising for low-temperature analog simulations. 

Within fermionic systems, we focused on the pure Hubbard model in this work. However, the microscopic model of cuprates is believed to possess a significant next-nearest-neighbor hopping $t'$ that explicitly breaks particle-hole symmetry and can engender distinct Hall anomalies.
These effects could also be studied using quantum simulators, which may be able to introduce a tunable $t'$ hopping, in order to understand the role of $t'$ in driving Fermi-surface reconstruction near the antinodal points \cite{pelz2026quantum}. 
Likewise, exploring how the quasiparticle transitions intertwine with low-temperature spiral, stripe or superconducting orders would be an exciting direction of research. 
It would also be interesting to compare the protocols discussed here with the doping-dependent Seebeck coefficient, which shows a sign change consistent with Fermi surface reconstruction in the Fermi-Hubbard model in the intermediate-coupling regime \cite{roy2026finite}.
Beyond ultracold atoms, these protocols could be adapted for solid-state settings using local magnetic noise spectroscopy \cite{rovny_nanoscale_2024,Curtis.2025}, paving the way for nanoscale measurements of non-Gaussian current fluctuations.


\pheader{Acknowledgments} We acknowledge insightful discussions with Alex Gomez Salvador, Roberta Citro, Thierry Giamarchi, Fabian Grusdt, Annabelle Bohrdt, Monika Aidelsburger, Waseem Bakr, Steven Kivelson, Dima Abanin, Trond Andersen, Sarang Gopalakrishnan and Assa Auerbach. The ETH group acknowledges funding from the SNSF project 200021\textunderscore212899, the SNSF Sinergia grant CRSII--222792 and the Swiss State Secretariat for Education, Research and Innovation (contract number UeM019-1).
U.B. is grateful for the financial support of the IBM Quantum Researcher Program.

\bibliography{bibliography.clean}





\onecolumngrid
\appendix
\newpage
\begin{center}
	\textbf{\Large Supplementary Information}
\end{center}
\normalsize

\setcounter{equation}{0}
\setcounter{figure}{0}
\setcounter{table}{0}
\renewcommand{\thefigure}{S\arabic{figure}}
\makeatletter
\setlength\tabcolsep{10pt}
\setcounter{secnumdepth}{2}

\section{Numerical Methods: Protocols}

We study a M-leg Fermi–Hubbard ladder using matrix product state methods, combining density matrix renormalization group (DMRG) for ground-state calculations and time dependent variational principle (TDVP) for real-time evolution, implemented in the ITensor/ITensorMPS framework. The system is defined on a lattice with sites $s=(m,l)$, where $m \in \{1,2,\dots,M\}$ labels the leg and $l \in \{1,\ldots,L\}$ denotes the rung position along the ladder. All protocols are implemented using the ITensors Julia libraries \cite{Fishman2022ITensor, Fishman2022ITensorRelease}.
Below, we first describe the the equilibrium protocol used to measure the quantum Hall coefficient and the subsystem density skewness of the many-body ground state, followed by the out-of-equilibrium nonlinear current response protocol.

\subsection{The Dimensionless Hall Constant ($r_H$)}
To set the stage for numerical calculations, we first show the analytic calculation of the Hall coefficient for noninteracting lattice fermions.

\subsubsection{Hall coefficient of non-interacting fermions}
Consider a 2D system of spinless fermions with charge $e$ on a square lattice. Setting the lattice constant $a = 1$, the nearest-neighbor tight-binding energy dispersion is
\begin{equation}
E(k_x, k_y) = -2t (\cos k_x + \cos k_y).
\end{equation}
The filling fraction $\nu \in [0,1]$ is defined by the integral over the filled states in the Brillouin zone ($[-\pi, \pi]^2$):
\begin{equation}
\nu = \int \frac{d^2k}{(2\pi)^2} \Theta(E_F - E(k_x, k_y)).
\end{equation}
The Jones-Zener expansion in the relaxation-time approximation of the Boltzmann equation can be used to find the longitudinal ($\sigma_{xx}$) and transverse ($\sigma_{xy}$) conductivities at $T=0$ \cite{PhysRev.56.93}. For the longitudinal case, we find
\begin{align}
\sigma_{xx} &= - \frac{e^2 \tau}{2\hbar^2} \int \frac{d^2k}{(2\pi)^2} \Theta(E_F - E) E(k_x, k_y) = - \frac{e^2 \tau}{2\hbar^2} \mathcal{E}(\nu),\label{eq:app:rHsigmaxx}
\end{align}
where $\mathcal{E}(\nu)$ is defined through the integral of the energy dispersion $E(k_x,k_y)$ in Eq. \eqref{eq:app:rHsigmaxx}.
For the transverse conductivity we obtain
\begin{align}
\sigma_{xy} &= \frac{e^3 B \tau^2}{\hbar^4} \int \frac{d^2k}{(2\pi)^2} \Theta(E_F - E) (\partial_{k_x}^2 E) (\partial_{k_y}^2 E) = \frac{4 e^3 B \tau^2 t^2}{\hbar^4} \Phi(\nu).
\end{align}
In the last equality, we substituted the tight-binding second derivatives of energy $(2t \cos k_x)(2t \cos k_y) = 4t^2 \cos k_x \cos k_y$ and defined the dimensionless integral over the filled states $\Phi(\nu)$:
\begin{equation}
\Phi(\nu) = \int \frac{d^2k}{(2\pi)^2} \Theta(E_F - E) \cos k_x \cos k_y.
\end{equation}

\paragraph{The Hall Coefficient}

In a standard Hall measurement ($J_y = 0$), the Hall coefficient is defined as $R_H = E_y / (J_x B)$. In terms of the conductivity tensor components (assuming weak fields where $\sigma_{xy} \ll \sigma_{xx}$), this is:
\begin{equation}
R_H = \frac{\sigma_{xy}}{B \sigma_{xx}^2}.
\end{equation}
Substituting our exact derivations for $\sigma_{xx}$ and $\sigma_{xy}$, we find the Hall coefficient
\begin{equation}
R_H(\nu) = \frac{ \frac{4 e^3 B \tau^2 t^2}{\hbar^4} \Phi(\nu) }{ B \left( - \frac{e^2 \tau}{2\hbar^2} \mathcal{E}(\nu) \right)^2 } = \frac{ \frac{4 e^3 \tau^2 t^2}{\hbar^4} \Phi(\nu) }{ \frac{e^4 \tau^2}{4\hbar^4} \mathcal{E}^2(\nu) }=
R_H(\nu) = \frac{16 t^2}{e} \frac{\Phi(\nu)}{[\mathcal{E}(\nu)]^2}.
\end{equation}

\begin{figure}[t]
        \centering
        \includegraphics[width = 0.5 \linewidth]{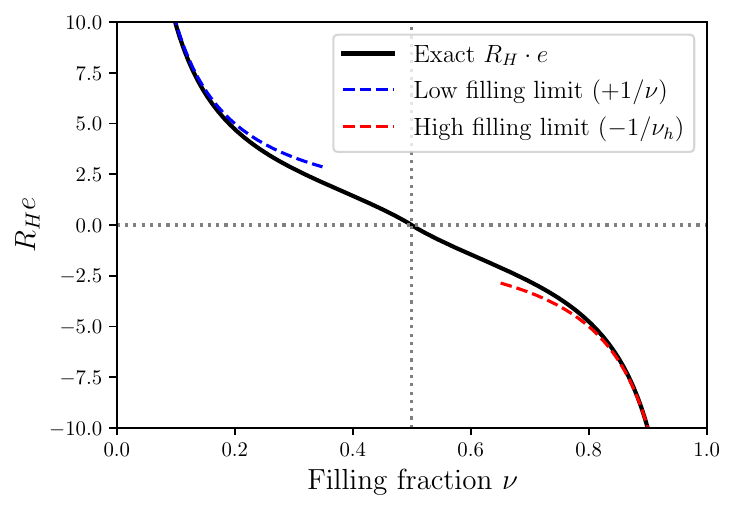}
        \caption{Dimensionless Hall coefficient of non-interacting fermions with nearest-neighbour hopping on a square lattice. The vertical dashed line corresponds to half-filling.}
        \label{fig:hall_noninteracting} 
\end{figure}

\paragraph{Filling Dependence of the Hall coefficient}

The full filling dependence of the Hall coefficient is shown in Figure \ref{fig:hall_noninteracting}. There are three interesting regimes:

\begin{itemize}
    \item \textbf{Empty Band Limit ($\nu \to 0$):} The filled states are tightly packed around $k = (0,0)$. Here, $\cos k_x \cos k_y \approx 1$, meaning $\Phi(\nu) \approx \nu$. The energy per particle is minimal, $E \approx -4t$, so that $\mathcal{E}(\nu) \approx -4t \nu$. The result
    \begin{equation}
    R_H \to \frac{16 t^2}{e} \frac{\nu}{16 t^2 \nu^2} = +\frac{1}{e \nu}
    \end{equation}
    matches the classical Drude result.

    \item \textbf{Full Band Limit ($\nu \to 1$):} Let $\nu_h = 1 - \nu$ be the hole concentration. The empty states are packed near $k = (\pm\pi, \pm\pi)$, where $\cos k_x \cos k_y \approx 1$ and $E \approx +4t$. Integrating over the Brillouin zone and subtracting the holes maps the integrals to $\Phi(\nu) \approx -\nu_h$ and $\mathcal{E}(\nu) \approx -4t \nu_h$. Again,
    \begin{equation}
    R_H \to \frac{16 t^2}{e} \frac{-\nu_h}{16 t^2 \nu_h^2} = -\frac{1}{e \nu_h}
    \end{equation}
    recovers the classical Drude result with the expected sign for oppositely charged holes.

    \item \textbf{Half Filling ($\nu = 0.5$):} At half filling, the fermi energy is zero; $E_F = 0$. The filled states occupy a diamond bounded by $|k_x| + |k_y| \le \pi$. Due to particle-hole symmetry, $\Phi(0.5)=0$. Consequently,
    \begin{equation}
    R_H(0.5) = 0.
    \end{equation}
\end{itemize}

\subsubsection{Hall coefficient of strongly-interacting fermions}

The Hall constant $R_H(T)$ of doped Mott insulators such as the cuprates exhibits
an anomalous dependence on both doping and temperature that has resisted a comprehensive theoretical description. Here we review and discuss our tensor-network-based
implementation of the approach introduced by Prelov\v{s}ek \emph{et al.}~\cite{prelovsek_hall_1999}.
We consider ladder geometries with periodic boundary conditions (PBC) along the
leg direction and open boundary conditions (OBC) along the transverse (rung)
direction. In this geometry, the DC transport response at $T=0$ can be extracted
directly from the \emph{ground-state energy}.
One must have three couplings —
an Aharonov--Bohm twist $\theta$, a transverse electric field (potential ramp)
$\Delta$, and a physical orbital flux $\varphi$ — as explicit parameters of the
Hamiltonian, and differentiate the ground-state energy $E(\theta,\Delta,\varphi)$
with respect to all three. We provide more details on the theory and numerical calculations below. The three ingredients that are included in addition to bare hopping are:

\begin{enumerate}
\item A synthetic Aharonov--Bohm flux $\Phi$ threaded through the periodic $x$
direction, which shifts every $x$-bond hopping phase by $\theta = e_0 \Phi / L$, where we take $e_0=+1$ as a convenient choice of charge unit, treating a single atomic count as a unit of charge.
This field exists purely so that the persistent/Drude current in the ring can be read off from $\partial E/\partial\theta$ via the Feynman--Hellmann theorem.
\item A uniform perpendicular magnetic field $B$, introduced via the Peierls
substitution. In the Landau gauge $\mathbf A = B(-y,0)$, only the $x$-direction
(``leg'') hoppings acquire a phase $\varphi = e_0 B$ (with $\hbar=1$).
\item A uniform transverse electric field $\mathcal E_y = \mathcal E \neq 0$ to act as the Hall field in equilibrium that counteracts the Hall current.
\end{enumerate}

With $m=1,\dots,M$ labeling legs, $i=1,\dots,L$ labeling rungs, $s=\uparrow,\downarrow$ labeling spin, and $\bar m \equiv (M+1)/2$ the center leg, the complete Hamiltonian is
\begin{align}
H &= H_x + H_y + H_\Delta + H_{\rm int}, \label{eq:H}\\[4pt]
H_x &= -t \sum_{m=1}^{M}\sum_{i,s} e^{\,i[(m-\bar m)\varphi - \theta]}\,
        c^\dagger_{m,i+1,s} c_{m,i,s} \;+\; \text{h.c.}, \label{eq:Hx}\\[4pt]
H_y &= -t' \sum_{m=1}^{M-1}\sum_{i,s} c^\dagger_{m+1,i,s} c_{m,i,s} \;+\; \text{h.c.}, \label{eq:Hy}\\[4pt]
H_\Delta &= \Delta \sum_{i,m} (m-\bar m)\, n_{m,i}, \label{eq:HDelta}\\[4pt]
H_{\rm int} & = U \sum_{m,i} n_{m,i,\uparrow}\, n_{m,i,\downarrow},
\label{eq:Hubbard}
\end{align}
where $\Delta = e_0 \mathcal E$ is the transverse potential ramp conjugate to the
Hall (polarization) field, $t$ is the leg (PBC, current-carrying) hopping, $t'$ is
the rung (OBC) hopping (here $t' = t$), and $H_{\rm int}$ is the Hubbard interaction.

The ground-state energy $E(\theta,\Delta,\varphi)$ is the main object of interest. Because
$\theta$ and $\Delta$ enter $H$ linearly (as coefficients of the current and dipole
operators respectively), the Hellmann--Feynman theorem gives the ground-state
electric current density $j=j_x$ and the transverse polarization density
$P = P_y$ directly as energy derivatives:
\begin{equation}
j = \frac{e_0}{V} \frac{\partial E}{\partial \theta}, \qquad
P = -\frac{e_0}{V} \frac{\partial E}{\partial \Delta},
\label{eq:FH}
\end{equation}
with $V = LM$ the number of lattice sites.

At $\varphi = 0$ (no physical field), one first looks for the \emph{equilibrium},
non-polar ground state, i.e., the $(\theta_0, \Delta_0)$ at which $j=0$ and $P=0$.
By reflection symmetry about the center of the ladder,
$m-\bar m \rightarrow -(m-\bar m)$,
the ground-state energy at zero magnetic field is an even function of the transverse
potential $\Delta$,
$E(\theta,\Delta,0)=E(\theta,-\Delta,0)$. Therefore, the equilibrium polarization vanishes at
$\Delta_0=0$. However, on a \emph{finite} ribbon, with PBC wrapping along the leg direction and OBC along the short transverse direction to the leg, minimizing $E(\theta,0,0)$ over $\theta$
generically produces $\theta_0 \neq 0$: a finite system can support a nonzero persistent current at ``zero field" purely as a finite-size effect, and this$\theta_0$ is expected to vanish only in the thermodynamic limit $L\to\infty$.
Using the correct $\theta_0$ as the expansion point is essential; expanding around
$\theta=0$ instead would contaminate the extracted response with this spurious finite-size current.

To model the Hall effect, one now turns on a small physical flux $\varphi\neq0$ and a
small twist increment $\tilde\theta$ away from $\theta_0$ (which sources a small
current $j\neq0$ through Eq.~\eqref{eq:FH}), and finds the required transverse field $\Delta$
to keep the system \emph{non-polar}, $P=0$ — i.e., the open-circuit Hall condition. This defines $\Delta(\tilde\theta,\varphi)$, and $R_H$ is then read off
from the ratio of the induced field to the current and the applied flux,
\begin{equation}
R_H \equiv -\frac{\mathcal E}{jB} = -\frac{\Delta}{j\varphi}.
\label{eq:RHdef}
\end{equation}

A Taylor expansion of $E(\theta_0+\tilde\theta,\Delta,\varphi)$ to third order,
using (i) the symmetry of the current operator under $\Delta\to-\Delta$ and (ii) the
antisymmetry of the diamagnetic (electric field induced) current and polarization operators under
$\varphi \to -\varphi$ (equivalently $\tilde\theta\to-\tilde\theta$ at fixed sign
conventions), eliminates most cross terms and leaves
\begin{equation}
E = E^0 + \tfrac12 E^0_{\theta\theta}\tilde\theta^2
        + \tfrac12 E^0_{\Delta\Delta}\Delta^2
        + E^0_{\Delta\varphi}\Delta\varphi
        + E^0_{\theta\Delta\varphi}\,\tilde\theta\,\Delta\,\varphi
        + \tfrac12 E^0_{\theta\Delta\Delta}\,\tilde\theta\,\Delta^2
        + \tfrac16 E^0_{\theta\theta\theta}\,\tilde\theta^3 + \cdots,
\label{eq:Etaylor}
\end{equation}
where the superscript $0$ denotes a derivative evaluated at the reference point
$(\theta_0,0,0)$. From Eqs.~\eqref{eq:FH} and \eqref{eq:Etaylor}, to leading order
the induced current is
\begin{equation}
j = \frac{e_0}{N} E^0_{\theta\theta}\,\tilde\theta,
\label{eq:jlead}
\end{equation}
while the zero-polarization condition $P=0$ fixes $\Delta$ order by order:
\begin{equation}
E^0_{\Delta\Delta}\Delta + E^0_{\Delta\varphi}\varphi
+ E^0_{\theta\Delta\varphi}\,\tilde\theta\,\varphi
+ E^0_{\theta\Delta\Delta}\,\tilde\theta\,\Delta = 0.
\label{eq:Pzero}
\end{equation}
Keeping terms linear in $\varphi$ and $\tilde\theta$,
\begin{equation}
\Delta = \Delta_\varphi + \Delta_j
= -\frac{E^0_{\Delta\varphi}}{E^0_{\Delta\Delta}}\,\varphi
  -\frac{\tilde E^0_{\theta\Delta\varphi}}{E^0_{\Delta\Delta}}\,\varphi\,\tilde\theta,
\qquad
\tilde E^0_{\theta\Delta\varphi} \equiv E^0_{\theta\Delta\varphi}
- \frac{E^0_{\Delta\varphi}\,E^0_{\theta\Delta\Delta}}{E^0_{\Delta\Delta}} .
\label{eq:Deltasplit}
\end{equation}
Here, $\Delta_\varphi$ is the equilibrium transverse imbalance required to compensate the polarization induced by the orbital flux $\varphi$ in the absence of a longitudinal current. The additional contribution $\Delta_j$, which is linear in both $\varphi$ and the current-generating parameter $\tilde\theta$, is the Hall voltage of interest.

Inserting $\Delta_j$ from Eq.~\eqref{eq:Deltasplit} and $j$ from
Eq.~\eqref{eq:jlead} into the definition Eq.~\eqref{eq:RHdef} gives the central result \cite{prelovsek_hall_1999},
\begin{equation}
R_H = \frac{V\, \tilde E^0_{\theta\Delta\varphi}}
{e_0\, E^0_{\Delta\Delta}\, E^0_{\theta\theta}}
\label{eq:RHmain}
\end{equation}
with $\tilde E^0_{\theta\Delta\varphi}$ defined in Eq.~\eqref{eq:Deltasplit}. Two
practical advantages are emphasized in the original paper: (i) only the ground-state
energy is needed instead of matrix elements to excited states or dynamical
susceptibilities; (ii) the zero-current, zero-polarization reference state of a
finite system is unambiguous and easy to locate numerically (a 1D minimization over
$\theta$, which we use later).

The dimensionless combination that we plot (and compute) in the main text is
\begin{equation}
r_H \equiv e_0 R_H / V = \frac{\tilde E^0_{\theta\Delta\varphi}}
{E^0_{\Delta\Delta}\,E^0_{\theta\theta}} .
\label{eq:rHdimless}
\end{equation}

\subsubsection{Numerical implementation: DMRG}
The implementation described in this section is based on the density-matrix
renormalization group (DMRG), using the Julia ITensors ecosystem to compute the
ground-state energies $E$ needed to evaluate the finite-difference derivatives
in Eq.~\eqref{eq:Deltasplit}. The overall procedure is as follows: we first
locate $\theta_0$, then evaluate the ground-state energy on a 
$3\times3\times3$ grid of parameters, use finite differences to obtain the
required derivatives, and finally calculate $r_H$ from Eq.~\eqref{eq:RHmain} and Eq.~\eqref{eq:rHdimless}
using the individual energies $E(\theta,\Delta,\varphi)$.

The DMRG
calculation additionally requires choosing a \emph{one-dimensional ordering} of
those sites along the matrix product state (MPS) chain, since every operator's
range in that ordering directly affects how hard DMRG has to work to represent it.
The implementation uses a \emph{leg-major} site ordering, which minimizes the range of the periodic bonds along the leg direction while restricting the longer-range interactions to the comparatively few rung couplings. Because $r_H$ ultimately comes from finite differences of $E$ up to \emph{third}
order, the sweep schedule is deliberately more
aggressive than a typical single-ground-state DMRG run would need. Bond dimension is
ramped from maximum dimenion $50$ up to $10000$ with a singular value truncation cutoff being $10^{-8}$ and absolute energy tolerance being $10^{-7}$.

\begin{figure}[t]
\centering
\begin{minipage}[t]{0.48\linewidth}
    \centering
    \includegraphics[width=\linewidth]{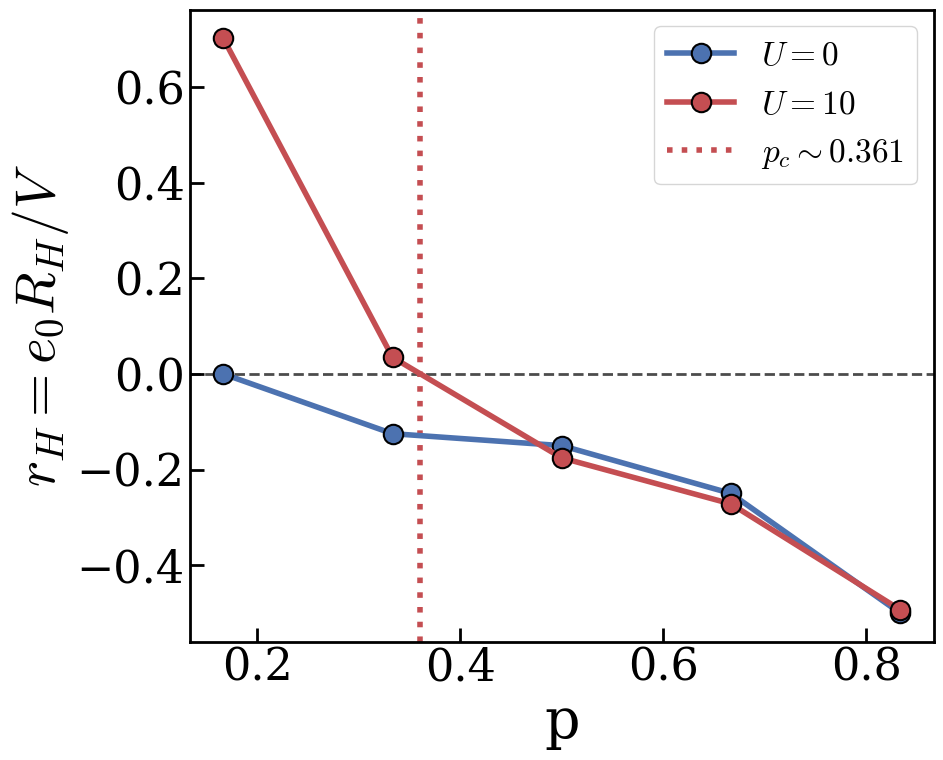}
\end{minipage}
\hfill
\begin{minipage}[t]{0.48\linewidth}
    \centering
    \includegraphics[width=\linewidth]{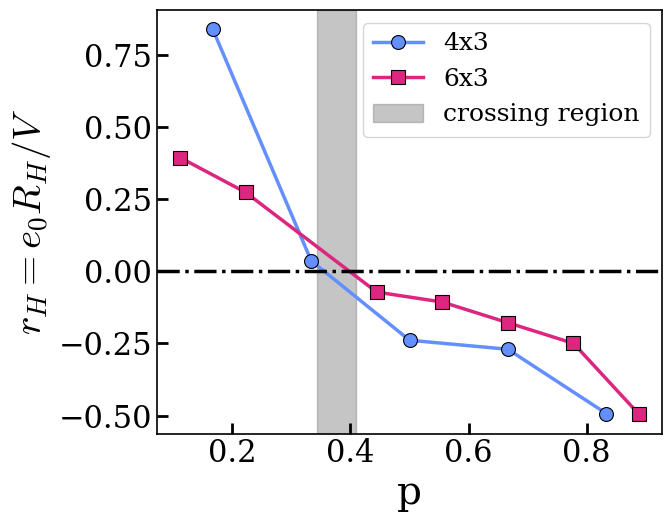}
\end{minipage}

\caption{
(a) Dimensionless Hall coefficient $r_H=e_0R_H/V$ as a function of hole doping for a $4\times3$ Hubbard ladder. The non-interacting system ($U=0$) retains the same (electron-like) sign of $r_H$ throughout the studied doping range and serves as a reference. In contrast, the strongly interacting system ($U/t_0=10$) undergoes a sign reversal at the characteristic doping $\delta_c\simeq0.36$ (vertical dashed line), signaling an interaction-driven anomaly in the Hall response.
(b) Finite-size comparison of the dimensionless Hall coefficient for $4\times3$ and $6\times3$ Hubbard ladders at $U/t_0=10$. Both system sizes exhibit a Hall sign reversal within the range $\delta_c\simeq0.35$--$0.41$, indicating only weak finite-size dependence of the characteristic doping and demonstrating that the interaction-driven Hall sign change is robust against system size.
}
\label{fig:hall_comparison}
\end{figure}

At $(\Delta,\varphi)=(0,0)$, the code evaluates $E(\theta,0,0)$ by DMRG at
dense equally spaced points over one flux quantum
$\theta \in [0, 2\pi/L)$. The finite-size equilibrium twist $\theta_0$ must be located
before any derivative is taken, since expanding around $\theta=0$ instead would
contaminate every downstream quantity with the ladder's spurious finite-size
persistent current. Around $(\theta_0, 0, 0)$, ground-state energies are evaluated by DMRG on all
$3^3 = 27$ points of a grid, 
\begin{equation}
(\theta,\Delta,\varphi) = \bigl(\theta_0 + a\,h_\theta,\; b\,h_\Delta,\; c\,h_\varphi\bigr),
\qquad a,b,c \in \{-1,0,1\},
\end{equation}
with step sizes $h_\theta=h_\Delta=h_\varphi=0.02$ by default. Writing
$f(a,b,c) \equiv E(\theta_0+a h_\theta,\, b h_\Delta,\, c h_\varphi)$, the five
derivatives needed for Eq.~\eqref{eq:RHmain} are estimated by standard
central finite-difference formulas:

\begin{align}
E^0_{\theta\theta} &\approx
\frac{f(1,0,0)-2f(0,0,0)+f(-1,0,0)}{h_\theta^2}, \\[2pt]
E^0_{\Delta\Delta} &\approx
\frac{f(0,1,0)-2f(0,0,0)+f(0,-1,0)}{h_\Delta^2}, \\[2pt]
E^0_{\Delta\varphi} &\approx
\frac{f(0,1,1)-f(0,1,-1)-f(0,-1,1)+f(0,-1,-1)}
{4\,h_\Delta h_\varphi}, \\[2pt]
E^0_{\theta\Delta\varphi} &\approx
\frac{
f(1,1,1)-f(1,1,-1)-f(1,-1,1)+f(1,-1,-1)}
{8\,h_\theta h_\Delta h_\varphi}
\nonumber\\
&\qquad
+\frac{
-f(-1,1,1)+f(-1,1,-1)+f(-1,-1,1)-f(-1,-1,-1)}
{8\,h_\theta h_\Delta h_\varphi}, \\[2pt]
E^0_{\theta\Delta\Delta} &\approx
\frac{
\bigl[f(1,1,0)-2f(1,0,0)+f(1,-1,0)\bigr]
-\bigl[f(-1,1,0)-2f(-1,0,0)+f(-1,-1,0)\bigr]}
{2\,h_\theta h_\Delta^2}.
\end{align}
 We assemble these to form $R_H$ as defined in Eq.\eqref{eq:RHmain} to finally obtain $r_H$ in Eq.\eqref{eq:rHdimless} which has been plotted in the main text. We show in Figure \ref{fig:hall_comparison}(a) the dimensionless Hall coefficient $r_H$ for a $4\times3$ ladder to compare the sign change between the non-interacting ($U=0$) and strongly interacting ($U=10$) regimes. In the non-interacting case, $r_H$ does not change sign over the accessible doping range, consistent with electron-like quasiparticles throughout. In contrast, the strongly interacting system exhibits a sign reversal at a critical doping $\delta_c \sim 0.36$, indicating a change in the nature of quasiparticles induced by the onsite repulsion. We also show in Fig.~\ref{fig:hall_comparison}(b) the finite-size effects on the doping range over which $r_H$ changes sign at $U/t_0=10$, by comparing ladders of sizes $4\times3$ and $6\times3$.

\subsection{Skewness-like density cumulant}

We first compute the third cumulant (skewness) of particle number fluctuations within a spatial subsystem $A$, defined as the connected third central moment
\begin{equation}
\langle (N_A - \langle N_A \rangle)^3 \rangle,
\end{equation}
where $N_A = \sum_{i \in A} n_i$ denotes the total particle number in the subsystem, $i$ is a two-dimensional spatial index marking the lattice sites and $n_i \equiv n_{i\uparrow} + n_{i\downarrow}$ is the local density operator. In our ladder geometry, the subsystem $A$ is chosen as a contiguous central block while excluding boundary sites along longitudinal and transverse directions. This minimizes edge effects and isolates bulk-like behavior, while maintaining a centrally located subsystem. 

Expanding the third central moment yields
\begin{equation}
\langle (N_A - \langle N_A \rangle)^3 \rangle
= \sum_{i,j,k \in A} \langle n_i n_j n_k \rangle
- 3 \langle N_A \rangle \sum_{i,j \in A} \langle n_i n_j \rangle
+ 2 \langle N_A \rangle^3,
\end{equation}
which separates contributions from three-point, two-point, and one-point correlators. The resulting skewness probes the asymmetry of the particle-number distribution and thus captures intrinsically non-Gaussian fluctuations. Importantly, while the total particle number of the full system is fixed, the subsystem $A$ is not a closed region and can exchange particles with its complement. Consequently, $N_A$ exhibits fluctuations similar to those of a subsystem coupled to an effective particle reservoir. 

\begin{figure}[t]
        \centering
        \includegraphics[width = 0.6 \linewidth]{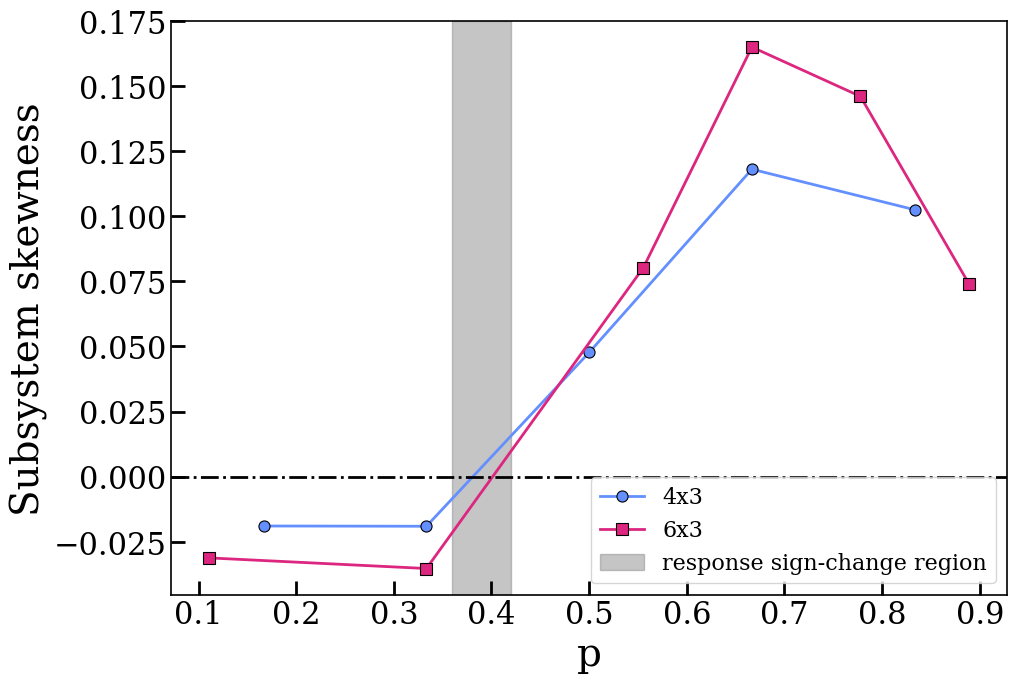}
        \caption{Subsystem skewness as a function of hole doping for two Fermi Hubbard ladder sizes, $4\times3$ and $6\times3$, in the interacting regime with $U/t_0 = 10$ where $t_0 = 1$ is the isotropic hopping strength. The skewness changes sign at nearly the same doping for both system sizes, with the zero crossing occurring in the range $\delta_c \simeq 0.36\text{--}0.42$ (indicated by the vertical grey shaded region). The weak system-size dependence of the crossing is consistent with the behavior of the dimensionless Hall constant $r_H$ shown in Fig.~\ref{fig:hall_comparison}, supporting a common characteristic hole doping range.
        }
        \label{fig:sub_skew_size} 
\end{figure}

We evaluate the third cumulant of $\langle N_A\rangle$ in equilibrium using the DMRG algorithm on the ground state of $4 \times 3$ upto $6 \times 3$ Fermi-Hubbard ladder, without incorporating any magnetic flux or onsite potential. For the ground-state calculations considered here, we employ open boundary conditions (OBC), which require a substantially smaller bond dimension than the subsequent calculations of $r_H$: the bond dimension is ramped from a maximum value of $50$ up to $5000$, with a singular-value truncation cutoff of $10^{-8}$ and an energy tolerance of $10^{-7}$. We plot the subsystem skewness in Fig.\ref{fig:sub_skew_size} for two different ladder sizes $4\times 3$ and $6 \times 3$ to show that the subsytem skewness changes sign consistently around the same hole filling $\delta_c \sim 0.36 - 0.42$, similar to that in Fig~\ref{fig:hall_comparison}.

After completion of the manuscript, private communications with two groups employing complementary tensor-network-based finite-temperature methods \cite{liyang_com} and NQS-based methods \cite{anikka_com} provided independent support for the sign change in the skewness that we observe.

\subsection{Non-Linear Response with Finite Momentum Onsite Potential Quench: Ground-State and Time Evolution}

We now describe the numerical protocol used to extract the nonlinear current response to a finite-momentum density quench. We first introduce the onsite potential quench and the Fermi-Hubbard ladder Hamiltonian without any additional fluxes, followed by the DMRG ground-state preparation and real-time TDVP evolution. We then describe how the second-order response at momentum $2q$ is properly isolated from the boundary-induced linear-response leakage, and finally discuss its expected short-time scaling and finite-size behavior.

\begin{figure}[htbp]
\centering
\begin{minipage}[t]{0.49\textwidth}
\centering
\includegraphics[width=\linewidth]{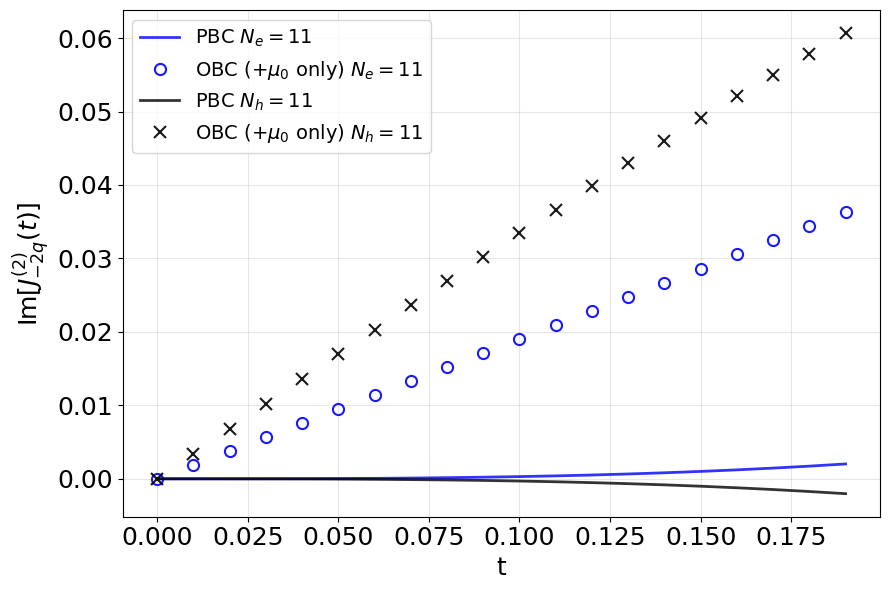}
\end{minipage}
\hfill
\begin{minipage}[t]{0.49\textwidth}
\centering
\includegraphics[width=\linewidth]{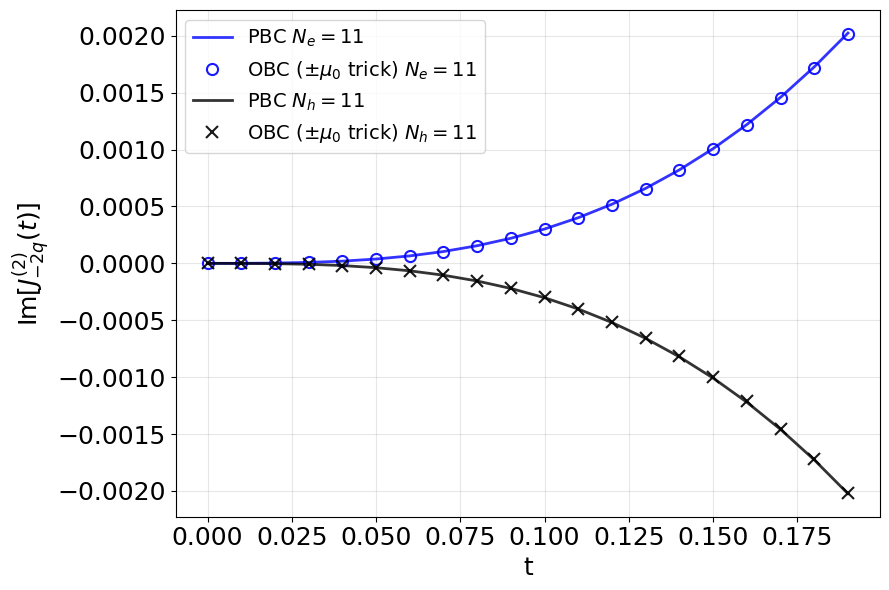}
\end{minipage}
\caption{Addressing the open-boundary leakage in the nonlinear ($p_x=2q$) current response. Both panels compare the non-linear response from $U = 0$ limit of Eq.~\eqref{eqn:H0} of dimensions $8\times4$ sites at fillings $N=11$ and $N_e=21$ ($N_h=11$), drive wavevector $q=2\pi/L$, and drive amplitude $\mu_0=1.0t_0$. (a) OBC (single $+\mu_0$ quench, no order isolation) vs.\ PBC. The $O(\mu_0^1)$ leakage from broken translational invariance visibly contaminates the nominally quadratic $p_x=2q$ channel. (b) OBC after applying $\pm\mu_0$ order isolation, $\big(\langle J_{2q}\rangle_{+\mu_0}+\langle J_{2q}\rangle_{-\mu_0}\big)/2$, compared with PBC. The unwanted $O(\mu_0^1)$ leakage is removed by field-reversal symmetrization, restoring agreement between the open- and periodic-boundary results.}
\label{fig:linear_nonlinear_comparison}
\end{figure}

We consider a quantum system subject to a weak, time-dependent external perturbation, with the total Hamiltonian
\begin{equation}
H(t)=H_0+H_1(t),
\end{equation}
where $H_0$ is the unperturbed Hamiltonian. We study a sudden density-potential quench switched on at $t=0$ with amplitude $\mu_0$,
\begin{equation}
H_1(t)=\mu_0\Theta(t)\sum_{x,y}\cos(qx)n_{x,y}\equiv\mu_0\Theta(t)\hat{V}_{pert},
\end{equation}
where $n_{x,y}=c^\dagger_{x,y}c_{x,y}$ is the local density operator and $\hat{V}_{pert}$ is the perturbing operator defined above. Numerically, we study a finite $M\times L$ Fermi--Hubbard ladder using matrix product state (MPS) techniques implemented with the ITensor library. The unperturbed Hamiltonian is
\begin{equation}
H_0=-t_{\rm leg}\sum_{\langle ij\rangle_{\rm leg},\sigma}(c^\dagger_{i\sigma}c_{j\sigma}+{\rm h.c.})-t_{\rm rung}\sum_{\langle ij\rangle_{\rm rung},\sigma}(c^\dagger_{i\sigma}c_{j\sigma}+{\rm h.c.})+U\sum_i n_{i\uparrow}n_{i\downarrow},
\label{eqn:H0}
\end{equation}
where $t_{\rm leg}$ and $t_{\rm rung}$ are the hopping amplitudes along the legs and rungs, respectively. Here open boundary conditions are used. The ground state at fixed $(N_\uparrow,N_\downarrow)$ is obtained using DMRG, starting from a random MPS with the desired quantum numbers and increasing the bond dimension up to $\chi_{\rm DMRG}=5000$, with truncation cutoff $10^{-8}$. The converged ground state is then evolved in real time following the quench using the two-site time-dependent variational principle (TDVP), with $\Delta t=0.04$, $\chi_{\rm TDVP}=3000 - 5000$, truncation cutoff $10^{-7}$, and Krylov dimension 7. The two-site formulation \cite{Haegeman16} permits the bond dimension to increase dynamically through SVD truncation, allowing the MPS to capture entanglement growth during the evolution.

\begin{figure}[hbtp]
\centering
\begin{minipage}[t]{0.48\textwidth}
\centering
\includegraphics[width=\linewidth]{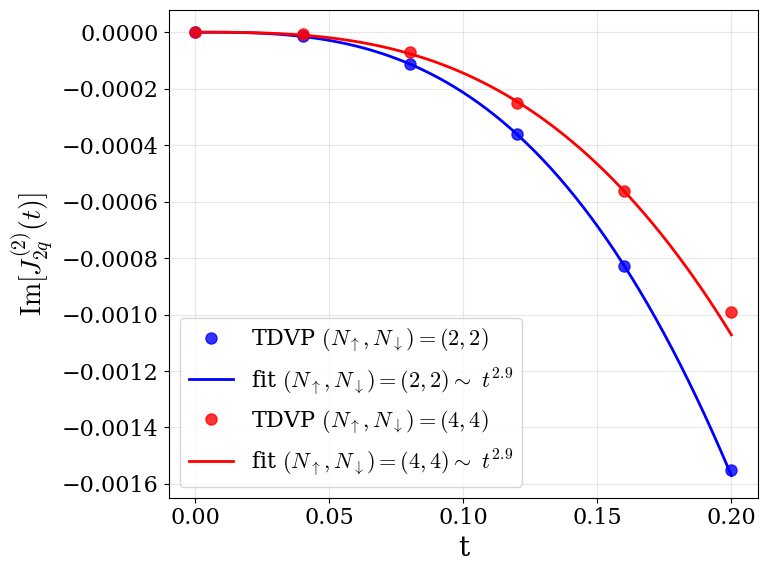}
\end{minipage}
\hfill
\begin{minipage}[t]{0.49\textwidth}
\centering
\includegraphics[width=\linewidth]{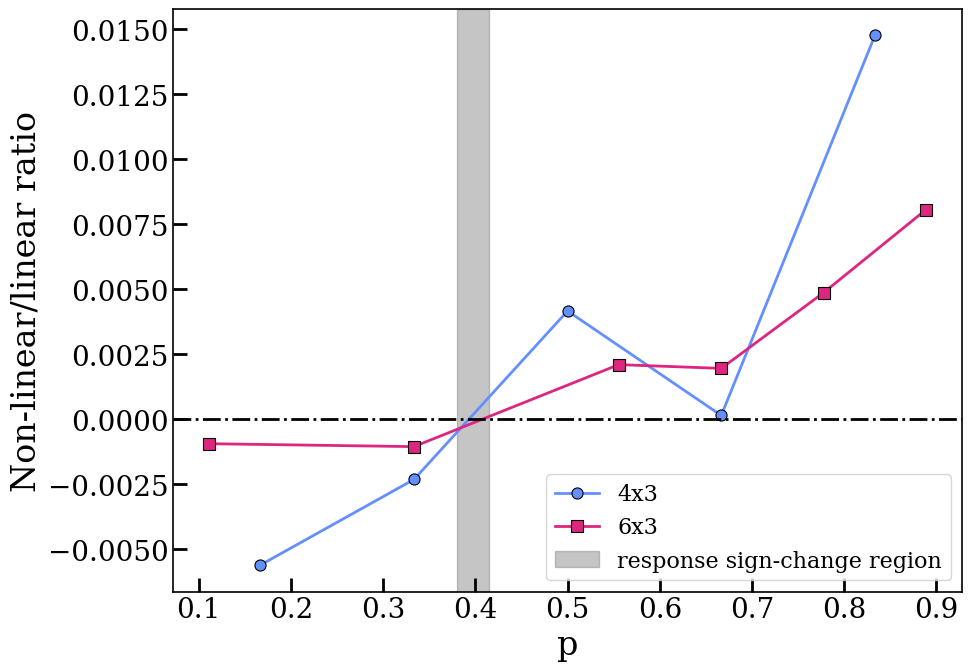}
\end{minipage}
\caption{(a) Nonlinear current response $\mathrm{Im}[\langle J_{2q}^{(2)}(t)\rangle]$ at momentum $2q$ for two particle-hole-complementary fillings, $(N_\uparrow,N_\downarrow)=(2,2)$ and $(4,4)$, on a $6\times3$ ladder. The TDVP results exhibit the expected short-time cubic scaling, $\mathrm{Im}[\langle J_{2q}^{(2)}(t)\rangle]\propto t^3$. Early-time power-law fits give exponents close to the predicted value, $\alpha\simeq2.9$. (b) Nonlinear-response ratio for two ladder sizes, $6\times3$ and $4\times3$, showing a crossing at hole doping $p_c\simeq0.38$--$0.42$.}
\label{fig:2d-nonlinearfit}
\end{figure}

To probe the dynamical response, we use $f_i=\cos(qx_i)$ with $q=2\pi/L_{\rm sites}$ and measure the momentum-resolved current
\begin{equation}
J_{p_x}(t)=\sum_j e^{-ip_xx_j}J_j(t).
\end{equation}
For open boundaries, broken translational invariance produces a linear-response contribution in the nominally nonlinear $p_x=2q$ channel. We therefore perform two independent quenches with $\mu_0=\pm V_{\rm quench}$ and isolate the second-order response through
\begin{equation}
J_{2q}^{(2)}(t)=\frac{1}{2}\left[J_{2q}(t,+V_{\rm quench})+J_{2q}(t,-V_{\rm quench})\right],
\end{equation}
which removes the leading $O(\mu_0)$ boundary-induced leakage while retaining the genuine $O(\mu_0^2)$ contribution.

In Fig.~\ref{fig:2d-nonlinearfit}(a), we verify that the nonlinear response exhibits the expected short-time cubic scaling. The $2q$ response arises from two insertions of the finite-momentum perturbation, while the corresponding time integrations in the second-order response generate a $t^3$ dependence. For a $6\times3$ ladder in the interacting regime $U/t_0=10$ with $t_{\rm leg}=t_{\rm rung}=t_0=1$, the imaginary part of the isolated response, $\mathrm{Im}[\langle J_{2q}^{(2)}(t)\rangle]$, scales as $t^3$ at short times for particle-hole-complementary fillings $(N_\uparrow,N_\downarrow)=(2,2)$ and $(4,4)$.  Early-time power-law fits give $\alpha\simeq2.9$ as can be observed in Fig.~\ref{fig:2d-nonlinearfit}(a). We further examine in Fig.~\ref{fig:2d-nonlinearfit}(b) finite-size scaling by comparing $6\times3$ and $4\times3$ ladders. The nonlinear-response ratio exhibits a stable crossing near hole doping $p_c\simeq0.38$--$0.42$, providing a consistent finite-size estimate of the characteristic doping scale associated with the nonlinear response.

\section{Connecting quench and spectroscopy protocols}

In this appendix, we provide an argument motivating why a sign change in the second-order response to a quench is related to a sign change in the steady-state response under periodic driving. We will build upon the connection between quenches and spectroscopy for linear response: see e.g. R. Kubo \emph{et al.} \cite{statistical_springer}.

Let the unperturbed Hamiltonian $\hat{H}_0$ have a stationary state $\hat{\rho}_0$. Couple a c-number field $f(t)$ to an observable $\hat{V}_{pert}$ and measure a second observable $\hat{A}$.
The linear response is governed by the retarded commutator
\begin{equation}\label{eq:app1-defs}
  \delta\langle \hat{A}\rangle^{(1)}(t)=\int_{-\infty}^{\infty}\!\dd t'\,\chi^{(1)}(t-t')f(t'),
  \qquad
  \chi^{(1)}(t)=i\,\Theta(t)\,\big\langle[\hat{A}(t),\hat{V}_{pert}]\big\rangle_0,
\end{equation}
and the two protocols of
interest differ only in the profile of $f$: a periodic drive measures $\chi^{(1)}(\omega)=\int_0^\infty\mathrm{d}t\,e^{i(\omega+i0^+)t}\chi^{(1)}(t)$, while a quench $f(t)=f_0\Theta(t)$ produces a trajectory $S^{(1)}(t)$ that is the time integral of the retarded response function:
\begin{equation}
  S^{(1)}(t)=\int_0^t\!\mathrm{d}s\;\chi^{(1)}(s).
\end{equation}
When $\hat{A}$ and $\hat{V}_{pert}$ carry the same parity under time reversal, such as in the density-density channel ($\hat{V}_{pert}=\hat{n}_{-\bm q}$, $\hat{A}=n_{\bm q}$), one has $\langle[\hat{A}(0),\hat{V}_{pert}]\rangle_0=0$, and $\langle[\hat{A}(t),\hat{V}_{pert}]\rangle_0$ is odd in $t$. Consequently $S^{(1)}(t)$ is even in time. The quench trajectory at short times can be expanded as a series:
\begin{equation}\label{eq:app1-pair}
  S^{(1)}(t)=\sum_{k\ge0}\frac{(-1)^kM_{2k+1}}{(2k+2)!}\,t^{2k+2},
\end{equation}
with the moments
\begin{equation}\label{eq:app1-M}
  M_{2k+1}\equiv\frac{2}{\pi}\int_0^\infty\!\mathrm{d}\omega\;
  \omega^{2k+1}\operatorname{Im}\chi^{(1)}(\omega) = -\avg{\big[\mathrm{ad}_H^{2k+1}\hat{A},\;\hat{V}_{pert}\big]}\qquad \mathrm{ad}_H\hat{X}\equiv[\hat{H}_0,\hat{X}],
\end{equation}
determined by the imaginary part of the response function. Equivalently, as seen from the last equality in \eqref{eq:app1-pair}, the moments $M_n$ can be obtained from ground-state expectation values of nested commutators of the unperturbed Hamiltonian $\hat{H}_0$ with the observable $\hat{A}$ and the perturbation $\hat{V}_{pert}$.
To sum up, equations \eqref{eq:app1-pair} and \eqref{eq:app1-M} show that the early-time behavior after a quench maps to moments of the frequency-domain response function, and thus connect quenches and periodic driving protocols.
A similar argument can be made when the operators have opposite time-reversal parity and $\langle[\hat{A}(0),\hat{V}_{pert}]\rangle_0\neq0$. In that case, the early-time expansion contains only odd powers or time. As before, the coefficients of the series are related to moments of the function $\operatorname{Im}\chi^{(1)}(\omega)$.

For second-order response, the general relation between the perturbation and the observable is 
\begin{equation}\label{eq:Rdef}
  \delta\langle \hat{A}\rangle^{(2)}(t)
  =\int_0^{\infty}\!\!\dd\tau_1\!\int_0^{\infty}\!\!\dd\tau_2\;
   R(\tau_1,\tau_2)\,f(t-\tau_1)\,f(t-\tau_1-\tau_2),
\end{equation}
with the kernel
\begin{equation}\label{eq:Rkernel}
  R(\tau_1,\tau_2)=-\big\langle\big[[\hat{A}(\tau_1+\tau_2),
  \hat{V}_{pert}(\tau_2)],\hat{V}_{pert}(0)\big]\big\rangle_0,
\end{equation}
which is now a function of two time variables; $\tau_2$ for the delay between the first and second insertion of the perturbation, and $\tau_1$ for the time between the second perturbation and the measurement \cite{PhysRevB.110.245132}.
As for first-order response, the early-time dynamics after a quench is described by a series
\begin{equation}\label{eq:app1-pair}
  S^{(2)}(t)=\sum_{k\ge0}\frac{R_{k}}{(k+2)!}\,t^{k+2}.
\end{equation}
Whereas the coefficients $M_n$ are related to moments of $\operatorname{Im}\chi^{(1)}(\omega)$ in the linear response case, the coefficients $R_k$ are tied to moments of $\varrho({\omega_1, \omega_2}) = \iint\!\dd\tau_1\dd\tau_2\;R(\tau_1,\tau_2)\,
  e^{i\omega_1\tau_1+i\omega_2\tau_2}$, the second-order analogue of spectral density \cite{PhysRevB.110.245132}.
The moments $C_{ab}$ form a two-index family because the two time intervals $\tau_1$ and $\tau_2$ map to two frequencies $\omega_1$ and $\omega_2$:
\begin{equation}\label{eq:S2-Cab}
  C_{ab}=-(-1)^{a+b}\iint\!\dd\omega_1\dd\omega_2\;
  \omega_1^{\,a}\,(\omega_2-\omega_1)^{\,b}\;\varrho(\omega_1,\omega_2).
\end{equation}
In analogy to linear response, the moments can be further related to nested commutators:
\begin{equation}
C_{ab}\equiv\Big\langle\Big[\big[\mathrm{ad}_H^{\,a}\hat{A},\;\mathrm{ad}_H^{\,b}\hat{V}_{pert}\big],\;\hat{V}_{pert}\Big]\Big\rangle_0.
\end{equation}
At order $k$, the expansion coefficients $R_k$ in \eqref{eq:app1-pair} are a linear combination of coefficients $C_{ab}$ such that $a+b=k$.
It follows that sign changes of coefficients in the early-time response \eqref{eq:app1-pair} will occur concomitantly with sign changes in certain linear combinations of moments of $\varrho(\omega_1, \omega_2)$ and the corresponding ground-state expectation values of nested commutators. 
Therefore, either early-time dynamics or spectroscopy can be used to probe the sign of charge carriers in cold-atom quantum simulators.

The exact relation between the family $R_k$ and the moments of $\varrho(\omega_1, \omega_2)$, and their particular form for the Fermi-Hubbard model are the subject of an upcoming work by the authors.

\section{Second harmonic generation on a ladder of hard core bosons}

In this section, we derive the second-order response of a square ladder of hard core bosons subject to an external drive coupling to density modulations at a finite momentum and frequency. We start by writing out the Hamiltonian and introducing the Gutzwiller mean-field variational wavefunction that we use to obtain analytical results. By solving the equations of motion for the variational parameters perturbatively in the drive strength, we find the first- and second-order response in the density-density and density-current channels. We show the sign of the second-order response is sensitive to deviations from half-filling whereas the first-order response is symmetric under a particle-hole transformation. Lastly, we comment on different types of drive (running and standing waves) and the relevance of Umklapp scattering for experiments.

\subsection{Hamiltonian description, variational wavefunction and equations of motion}
The system Hamiltonian of hard core bosons on a square two-leg ladder is 
\begin{equation}
    \hat{H} = -t_\parallel\sum\limits_{n,l}\left(\hat{b}_{n,l}^\dag \hat{b}_{n,l+1} + \hat{b}_{n,l+1}^\dag \hat{b}_{n,l}\right) - t_\perp\sum\limits_l \left(\hat{b}_{1,l}^\dag \hat{b}_{2,l} + \hat{b}_{2,l}^\dag \hat{b}_{1,l}\right)-\mu\sum\limits_{n,l}\hat{b}^\dag_{n,l}\hat{b}_{n,l},
\end{equation}
where the index $n=1,2$ denotes the leg of the ladder and $l$ denotes the rung. We can map to a spin-$1/2$ problem by $\hat{S}^z=\hat{b}^\dag \hat{b}-\frac{1}{2}$ and $\hat{S}^+=\hat{b}^\dag$ to obtain
\begin{equation}
    \hat{H} = -t_\parallel\sum\limits_{n,l}\left(\hat{S}_{n,l}^+ \hat{S}_{n,l+1}^- + \hat{S}_{n,l+1}^+ \hat{S}_{n,l}^-\right) - t_\perp\sum\limits_l \left(\hat{S}_{1,l}^+ \hat{S}_{2,l}^- + \hat{S}_{2,l}^+ \hat{S}_{1,l}^-\right)-\mu\sum\limits_{n,l}\hat{S}^z_{n,l}.
\end{equation}
We will analyze the response to a longitudinal perturbation
\begin{equation}
    \delta \hat{H} =\sum\limits_{n,l}\delta V_{n,l}\hat{S}^z_{n,l},\label{eq:longitudinal_drive}
\end{equation}
with $\delta V_{n,l}=\delta V e^{i(ql - \omega t)}+h.c.$, which can be thought of as a monochromatic plane-wave component of a standing wave drive. From now on, we will focus on the response to the $(q,\omega)$ complex component of the drive -- the response to $(-q,-\omega)$ is its hermitian conjugate. We will work with the Gutzwiller mean-field wave function
\begin{equation}
    \ket{\psi_{MF}} = \prod\limits_{n,l}\left(\mathrm{sin}\left(\frac{\theta_{n,l}}{2}\right)e^{-i\frac{\varphi_{n,l}}{2}}\ket{\downarrow_{n,l}} + \mathrm{cos}\left(\frac{\theta_{n,l}}{2}\right)e^{i\frac{\varphi_{n,l}}{2}}\ket{\uparrow_{n,l}}\right).
\end{equation}
With the expectation values,
\begin{align}
    \langle \hat{S}^z\rangle = \frac{1}{2}\mathrm{cos}(\theta),\quad\quad
    \langle \hat{S}^+\rangle = \frac{1}{2}\mathrm{sin}(\theta)e^{-i\varphi},\quad\quad
    \langle \hat{S}^-\rangle = \frac{1}{2}\mathrm{sin}(\theta)e^{i\varphi},\quad\quad
    \langle i\frac{\partial}{\partial t}\rangle = -\sum\limits_{n,l}\mathrm{cos}(\theta_{n,l})\frac{\dot{\varphi}_{n,l}}{2},
\end{align}
we get the Lagrangian
\begin{align}
    L &=  \langle i\frac{\partial}{\partial t}\rangle- \langle \hat{H}\rangle\nonumber\\
    &=-\sum\limits_{n,l}\mathrm{cos}(\theta_{n,l})\frac{\dot{\varphi}_{n,l}}{2} + \frac{t_\parallel}{2} \sum\limits_{n,l}\mathrm{sin}(\theta_{n,l}) \mathrm{sin}(\theta_{n,l+1}) \mathrm{cos}(\varphi_{n,l+1}-  \varphi_{n,l}) + \nonumber\\
    &\quad+\frac{t_\perp}{2}\sum\limits_l \mathrm{sin}(\theta_{1,l}) \mathrm{sin}(\theta_{2,l}) \mathrm{cos}(\varphi_{1,l}-\varphi_{2,l}) + \frac{\mu}{2}\sum\limits_{n,l} \mathrm{cos}(\theta_{n,l})  - \frac{1}{2} \sum\limits_{n,l} \delta V_{n,l}\mathrm{cos}(\theta_{n,l}),
\end{align}
and from it the variational equations of motion
\begin{align}
    \mathrm{sin}(\theta_{n,l})\dot{\theta}_{n,l} &= t_\parallel\mathrm{sin}(\theta_{n,l+1})\mathrm{sin}(\theta_{n,l})\mathrm{sin}(\varphi_{n,l+1}-\varphi_{n,l}) - t_\parallel\mathrm{sin}(\theta_{n,l-1})\mathrm{sin}(\theta_{n,l})\mathrm{sin}(\varphi_{n,l}-\varphi_{n,l-1}) +\nonumber\\
    &\quad+ t_\perp\mathrm{sin}(\theta_{1,l})\mathrm{sin}(\theta_{2,l})\mathrm{sin}(\varphi_{1,l}-\varphi_{2,l})(\delta_{n,2}-\delta_{n,1}),\label{eq:eqm_phi}\\
    -\mathrm{sin}(\theta_{n,l}) \dot{\varphi}_{n,l} &= t_\parallel\mathrm{cos}(\theta_{n,l}) \mathrm{sin}(\theta_{n,l+1}) \mathrm{cos}(\varphi_{n,l+1}-\varphi_{n,l}) + t_\parallel\mathrm{cos}(\theta_{n,l})\mathrm{sin}(\theta_{n,l-1}) \mathrm{cos}(\varphi_{n,l-1}-\varphi_{n,l})+ \nonumber\\
    &\quad + t_\perp\mathrm{cos}(\theta_{1,l})\mathrm{sin}(\theta_{2,l}) \mathrm{cos}(\varphi_{1,l}-\varphi_{2,l}) \delta_{1,n} + t_\perp\mathrm{sin}(\theta_{1,l}) \mathrm{cos}(\theta_{2,l})\mathrm{cos}(\varphi_{1,l}-\varphi_{2,l})\delta_{2,n}+\nonumber\\
    &\quad+ \delta V_{n,l} \mathrm{sin}(\theta_{n,l}) -\mu\mathrm{sin} (\theta_{n,l}).\label{eq:eqm_theta}
\end{align}
In the translationally invariant ground state with $\delta V \equiv 0$, all the parameters are the same on all sites. The value of $\varphi_0$ is arbitrary and can be set to 0. The ground-state value of $\theta$ is determined by the filling through $\mathrm{cos}(\theta_0)=2\nu-1$ for the chemical potential $\mu = c_0 (2t_\parallel+ t_\perp)$. We see that $\mathrm{cos}(\theta_0)$ measures deviation from half filling and will flip sign under a particle-hole transformation.

On a lattice, the longitudinal and transverse currents and their expectation values are defined as
\begin{align}
    \hat{J}_\parallel(n,l) &= it_\parallel(\hat{S}^+_{n,l+1}\hat{S}^-_{n,l} - \hat{S}^+_{n,l}\hat{S}^-_{n,l+1}),\\
    J_\parallel(n,l)  &\equiv \langle \hat{J}_\parallel(n,l) \rangle = \frac{t_\parallel}{2}\mathrm{sin}(\theta_{n,l+1})\mathrm{sin}(\theta_{n,l})\mathrm{sin}(\varphi_{n,l+1}-\varphi_{n,l}),\\
    \hat{J}_\perp(l) &= it_\perp(\hat{S}_{1,l}^+\hat{S}_{2,l}^- - \hat{S}_{2,l}^+ \hat{S}_{1,l}^-),\\
    J_\perp(l) &\equiv \langle \hat{J}_\perp(l)\rangle= \frac{t_\perp}{2} \mathrm{sin}(\theta_{1,l})\mathrm{sin}(\theta_{2,l})\mathrm{sin}(\varphi_{1,l}-\varphi_{2,l}).
\end{align}
In general, we will use the shorthand $A=\langle \hat{A}\rangle$. The continuity equation demands $\frac{\mathrm{d}}{\mathrm{d}t}\rho(n,l) = J_\parallel(n,l-1) - J_\parallel(n,l) - (-1)^n J_\perp(l)$ at every site. In the ground state, all currents vanish. Since we are driving the ladder longitudinally, the transverse current will remain zero even after turning on the drive. To compute the response in density modulation and currents to an external drive, we expand around the equilibrium up to second order in perturbation 
\begin{align}
    &\theta_{n,l}(t) = \theta_0 + \delta\theta_{n}^{(1)} e^{i(ql - \omega t)} + \delta\theta_{n}^{(2)} e^{i(2ql - 2\omega t)},\\
    &\varphi_{n,l}(t) = \delta\varphi_{n}^{(1)}e^{i(ql - \omega t)} + \delta\varphi_{n}^{(2)}e^{i(2ql - 2\omega t)},
\end{align}
where $\theta_{n}^{(1)}$ and $\varphi_{n}^{(1)}$ will turn out to be proportional to $\delta V$ and $\theta_{n}^{(2)}$ and $\varphi_{n}^{(2)}$ will be proportional to $\delta V^2$. We will find useful the expansions
\begin{align}
    &\mathrm{cos}\left(\theta_0 + \delta\theta_{n,l}^{(1)}+ \delta \theta_{n,l}^{(2)}\right) \approx c_0 - s_0 \delta \theta_{n,l}^{(1)} - \frac{c_0}{2}(\delta \theta_{n,l}^{(1)})^2 - s_0 \delta \theta_{n,l}^{(2)},\\
    &\mathrm{sin}\left(\theta_0+ \delta\theta_{n,l}^{(1)}+ \delta \theta_{n,l}^{(2)}\right) \approx s_0 + c_0 \delta \theta_{n,l}^{(1)} - \frac{s_0}{2}( \delta \theta_{n,l}^{(1)})^2 + c_0\delta\theta_{n,l}^{(2)},
\end{align}
which are valid up to second order in perturbation.

\subsection{Linear response}
In this section, we show how to find the first-order response in density and current modulations to the external drive \eqref{eq:longitudinal_drive}. We plug the perturbative expansions of the variational parameters into \cref{eq:eqm_phi,eq:eqm_theta} to find, to first order
\begin{equation}
    M^{(1)}\begin{pmatrix}
        \delta\theta_1^{(1)} \\ \delta\theta_2^{(1)}\\ \delta\varphi_1^{(1)} \\ \delta\varphi_2^{(1)}
    \end{pmatrix} = \begin{pmatrix}
       0 \\ 0 \\  \delta Vs_0\\ \delta V s_0
    \end{pmatrix},\label{eq:first_order_mateq}
\end{equation}
with the matrix
\begin{equation}
    M^{(1)}=\begin{pmatrix}
        i\omega & 0 & 2t_\parallel s_0(\mathrm{cos}(q)-1)-t_\perp s_0 & t_\perp s_0\\
        0 & i\omega & t_\perp s_0 & 2t_\parallel s_0(\mathrm{cos}(q) -1)-t_\perp s_0\\
        -2t_\parallel c_0^2\mathrm{cos}(q) + 2t_\parallel + t_\perp & -t_\perp c_0^2& i\omega s_0&0\\
        -t_\perp c_0^2 &- 2t_\parallel c_0^2\mathrm{cos}(q) + 2t_\parallel + t_\perp & 0 &i\omega s_0
    \end{pmatrix},\label{dispersion_matrix_M1}
\end{equation}
where we use shorthand $c_0=\mathrm{cos}(\theta_0)$, $s_0=\mathrm{sin}(\theta_0)$. By setting the determinant of $M^{(1)}$ to zero, we find the two branches of collective excitations
\begin{align}
    \omega_{1+}^2(q) &= 2t_\parallel(1-\mathrm{cos}(q))\left( 2t_\parallel(1- c_0^2 \mathrm{cos}(q)) + t_\perp s_0^2\right)\label{eq:dispersion_+},\\
    \omega_{1-}^2(q) &= 2\left(t_\parallel(1-\mathrm{cos}(q))+t_\perp\right)\left( 2t_\parallel(1 -c_0^2\mathrm{cos}(q)) + (1+c_0^2) t_\perp\right).\label{eq:dispersion_-}
\end{align}
Both branches have a minimum at $q=0$ and nowhere else. Applying a particle-hole transformation does not affect the position of the resonances. Whereas \eqref{eq:dispersion_+} corresponds to particles on the two legs moving symmetrically, \eqref{eq:dispersion_-} corresponds to opposite oscillations of the two legs. We see the symmetric mode is gapless while the antisymmetric mode is gapped if $t_\perp\neq 0$. The gapless mode is the Goldstone mode of the broken $U(1)$ in-plane rotational symmetry. Note that the frequencies are real for all momenta, which means that within the variational space, we linearized around a ground state. The two modes coalesce when $t_\perp=0$, since we are then describing two decoupled chains. In the long-wavelength limit, the symmetric mode disperses as
\begin{equation}
    \lim_{q\to 0}\omega_{1+}(q) = |q|s_0 \sqrt{t_\parallel(2t_\parallel+t_\perp)}.
\end{equation}
Back to the equation \eqref{eq:first_order_mateq} for the driven system, we solve it to get
\begin{equation}
    \begin{pmatrix}
        \delta\theta^{(1)} \\ \delta \varphi^{(1)}
    \end{pmatrix} = \frac{\delta V}{\omega^2-\omega^2_{1+}(q)}\begin{pmatrix}
        2t_\parallel s_0(\mathrm{cos}(q) - 1)\\ -i\omega
    \end{pmatrix},
\end{equation}
where we omit the leg index because the solution is symmetric. We now want to look at the first-order Fourier component at $(q,\omega)$ of the density response $\rho(n,q) = \frac{1}{L}\sum\limits_{l} \rho(n,l) e^{-iql}$ with $\rho(n,l)=\langle\hat{b}_{n,l}^\dag \hat{b}_{n,l}\rangle$, which can be rewritten in terms of variational parameters as
\begin{align}
    \rho_1(n,q,\omega) &= -\frac{s_0}{2}\delta\theta_{n,l}^{(1)},
\end{align}
and is evaluated to
\begin{equation}
    \rho_1(n,q,\omega) =\frac{t_\parallel s_0^2 \delta V (1-\mathrm{cos}(q))}{\omega^2-\omega_{1+}^2(q)},
\end{equation}
which is invariant under the particle-hole transformation, since there are no odd powers of $c_0$. The response is the same on both legs,which is why we omit the leg index in the main text. To include damping, we shift all frequencies $\omega\rightarrow\omega+i\eta$. In addition to a density response, the periodic drive also induces a current response at first order. We compute the Fourier component at $(q,\omega)$ of the longitudinal current $J_{1\parallel}(n,q) = \frac{1}{L}\sum\limits_{l} J_{1\parallel}(n,l) e^{-iql}$, since that will be the only one with linear response. We find
\begin{equation}
    J_{1\parallel}(n,q,\omega) = \frac{t_\parallel s_0^2\omega\delta V}{2(\omega^2-\omega^2_{1+}(q))} \left(\mathrm{sin}(q) + i(1-\mathrm{cos}(q))\right),
\end{equation}
which is again invariant under the particle-hole transformation. In fact, the density and current responses are connected by means a continuity equation: 
\begin{equation}
     J_{1\parallel}(n,q,\omega) = \frac{i\omega}{1- e^{-iq}} \rho_1(n,q,\omega),
\end{equation}
which simplifies to 
\begin{equation}
     J_{1\parallel}(n,q,\omega) = \frac{\omega}{q} \rho_1(n,q,\omega)
\end{equation}
in the long-wavelength limit.

\subsection{Second-order response}
\begin{figure}[h!]
        \centering
        \includegraphics[width = 0.8 \linewidth]{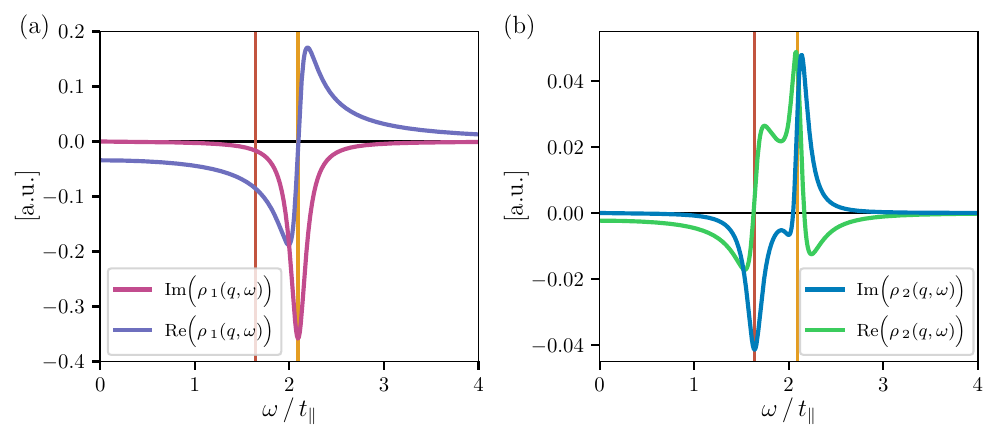}
        \caption{\textbf{Frequency dependence of the first- and second-order density modulations as a function of the drive frequency.} Data shown is for transverse hopping $t_\perp= 0.25t_\parallel$, damping rate $\eta=0.1t_\parallel$, driving strength $\delta V =0.2t_\parallel$ and filling $\nu=0.25$, at a fixed drive momentum $q=\pi/2$. Vertical lines denote the linear (orange) and nonlinear (brick red) resonance. (a) Imaginary and real part of the first-order density response. (b) Imaginary and real part of the second-order density response.}
        \label{fig:imre_app} 
\end{figure}

Now that we found the first-order response function, we want to calculate the response in density and current modulations at second order.
We expand equations of motion for a periodic monochromatic drive up to second order to find
\begin{align}
    M^{(2)}\begin{pmatrix}
        \delta\theta_1^{(2)} \\ \delta\theta_2^{(2)}\\ \delta\varphi_1^{(2)} \\ \delta\varphi_2^{(2)}
    \end{pmatrix} = \begin{pmatrix*}[l]
        2t_\parallel c_0\delta\theta_1^{(1)}\delta\varphi_1^{(1)} (\mathrm{cos}(q)-\mathrm{cos}(2q)) + t_\perp c_0 \delta\theta_2^{(1)} (\delta\varphi_1^{(1)} - \delta\varphi_2^{(1)})\\  
        2t_\parallel c_0\delta\theta_2^{(1)} \delta\varphi_2^{(1)} (\mathrm{cos}(q)-\mathrm{cos}(2q)) - t_\perp c_0 \delta\theta_1^{(1)} (\delta\varphi_1^{(1)} - \delta\varphi_2^{(1)})\\ 
        -t_\parallel c_0s_0(\delta\varphi_1^{(1)})^2\left( \mathrm{cos}(2q)- 2\mathrm{cos}(q) + 1 \right) -t_\parallel c_0s_0(\delta\theta_1^{(1)})^2 \left( \mathrm{cos}(2q) + 2\mathrm{cos}(q)\right) -\\
        \quad\quad\quad\quad-i\omega c_0\delta \theta_1^{(1)}\delta\varphi_1^{(1)} - \frac{t_\perp}{2}c_0s_0 \left(\left( \delta\varphi_1^{(1)} - \delta\varphi_2^{(1)}\right)^2 +  2\delta\theta_1^{(1)}  \delta\theta_2^{(1)}+ (\delta\theta_2^{(1)})^2 \right) +\delta V c_0\delta\theta_1^{(1)}\\
        -t_\parallel c_0s_0(\delta\varphi_2^{(1)})^2\left( \mathrm{cos}(2q)- 2\mathrm{cos}(q) + 1 \right) -t_\parallel c_0s_0(\delta\theta_2^{(1)})^2 \left( \mathrm{cos}(2q) + 2\mathrm{cos}(q)\right) -\\
        \quad\quad\quad\quad-i\omega c_0\delta \theta_2^{(1)} \delta\varphi_2^{(1)} - \frac{t_\perp}{2}c_0s_0 \left(\left( \delta\varphi_1^{(1)} - \delta\varphi_2^{(1)}\right)^2 +  2\delta\theta_1^{(1)}  \delta\theta_2^{(1)}+ (\delta\theta_1^{(1)})^2 \right) +\delta V c_0\delta\theta_2^{(1)}
    \end{pmatrix*},\label{eq:second_order_self2}
\end{align}
with the matrix $M^{(2)}$ equal to
\begin{equation}
    \begin{pmatrix}
        2i\omega & 0 & 2t_\parallel s_0(\mathrm{cos}(2q)-1)-t_\perp s_0 & t_\perp s_0\\
        0 & 2i\omega & t_\perp s_0 & 2t_\parallel s_0(\mathrm{cos}(2q) -1)-t_\perp s_0\\
        -2t_\parallel c_0^2\mathrm{cos}(2q) + 2t_\parallel + t_\perp & -t_\perp c_0^2 & 2i\omega s_0&0\\
        -t_\perp c_0^2 & -2t_\parallel c_0^2\mathrm{cos}(2q)+ 2t_\parallel + t_\perp & 0 & 2i\omega s_0
    \end{pmatrix} \label{dispersion_matrix_M2}.
\end{equation}
Notice the R.H.S. of \eqref{eq:second_order_self2} is proportional to $c_0$, which means the variational parameters $\theta^{(2)}_n$ and $\varphi^{(2)}_n$ will be as well. From the determinant of $M^{(2)}$, we can extract ''non-linear" resonances at 
\begin{align}
    \omega_{2+}^2(q) &= \frac{t_\parallel}{2}(1-\mathrm{cos}(2q))\left( 2t_\parallel(1- c_0^2 \mathrm{cos}(2q)) + t_\perp s_0^2\right)\label{dispersion_second+},\\
    \omega_{2-}^2(q) &= \frac{1}{2}\left(t_\parallel(1-\mathrm{cos}(2q))+t_\perp\right)\left( 2t_\parallel(1 -c_0^2\mathrm{cos}(2q)) + (1+c_0^2) t_\perp\right),\label{dispersion_second-}
\end{align}
which are again gapless and gapped, respectively. We observe the symmetric mode disperses the same as the linear resonance at low momenta but differs at higher $q$ where the lattice structure becomes relevant. We see that $\omega_{2\pm}(q)$ can we obtained from $\omega_{1\pm}(q)$ by substituting $q\rightarrow 2q$ and $\omega\rightarrow 2\omega$, which can be understood as two-wave mixing. Since the first-order response is symmetric across the legs, we can simplify the second-order equation of motion to
\begin{equation}
   M^{(2)}\begin{pmatrix}
        \delta\theta_1^{(2)} \\ \delta\theta_2^{(2)}\\ \delta\varphi_1^{(2)} \\ \delta\varphi_2^{(2)}
    \end{pmatrix} =  \begin{pmatrix*}[l]
       r_\theta^{(2)} \\ r_\theta^{(2)} \\ r_\varphi^{(2)} \\r_\varphi^{(2)}
    \end{pmatrix*}
\end{equation}
with
\begin{align}
    r_\theta^{(2)} &=2t_\parallel c_0\delta\theta^{(1)}\delta\varphi^{(1)} (\mathrm{cos}(q)-\mathrm{cos}(2q)),\\
    r_\varphi^{(2)} &= -t_\parallel c_0s_0(\delta\varphi^{(1)})^2\left( \mathrm{cos}(2q)- 2\mathrm{cos}(q) + 1 \right) -t_\parallel c_0s_0(\delta\theta^{(1)})^2 \left( \mathrm{cos}(2q) + 2\mathrm{cos}(q)\right) -\nonumber\\
    &\quad-i\omega c_0\delta \theta^{(1)}\delta\varphi^{(1)} - \frac{3t_\perp}{2}c_0s_0 (\delta\theta^{(1)})^2  +\delta V c_0\delta\theta^{(1)}.
\end{align}
The solution is found to be 
\begin{align}
    \begin{pmatrix}
        \delta\theta^{(2)} \\
        \delta\varphi^{(2)}
    \end{pmatrix} = \frac{1}{2 \left(\omega^2 - \omega_{2+}^2(q)\right)}\begin{pmatrix}
        t_\parallel r_\varphi^{(2)}(\mathrm{cos}(2q) - 1) - i\omega r_\theta^{(2)} \\
        \frac{1}{s_0}\left( t_\parallel r_\theta^{(2)} (1 - c_0^2\mathrm{cos}(2q)) - i\omega r_\varphi^{(2)} + \frac{t_\perp}{2}s_0^2 r_\theta^{(2)}\right)
    \end{pmatrix},
\end{align}
where we omitted the leg index for clarity. The expansion of density modulations at $(2q,2\omega)$ up to second order gives
\begin{align}
    \rho_2(n,q,\omega)= -\frac{c_0}{4}(\delta\theta_{n,l}^{(1)})^2 - \frac{s_0}{2}\delta\theta_{n,l}^{(2)},
\end{align}
and for longitudinal current up to second order
\begin{align}
    J_{2\parallel}(n,q,\omega) =\frac{t_\parallel}{2}\left( s_0^2\delta\varphi_n^{(2)} + c_0s_0\delta\theta_n^{(1)} \delta\varphi_n^{(1)} \right)(e^{i2q}-1),
\end{align}
which are both proportional to $c_0$ and therefore to the deviation from half-filling. Consequently, the second-order density modulations at $(2q,2\omega)$ and the second-order current response at $(2q,2\omega)$ will flip sign under a particle-hole transformation. The two observables are again connected by the continuity equation
\begin{equation}
    J_{2\parallel}(n,q,\omega)= \frac{i2\omega}{1- e^{-2iq}}\rho_2(n,2q,2\omega).
\end{equation}  
Figure \ref{fig:imre_app} shows the the first- and second-order density response as a function of the drive frequency. We see the real part of the nonlinear response function has a node at the nonlinear collective mode $\omega=\omega_{2+}(q)$, whereas the behavior near the first-order pole is more complex.

\subsection{Experimental considerations}
In an experiment, the drive will not be a complex-valued plane wave but rather a real observable -- for example, a running wave;  $\delta \hat{H} = \sum\limits_{n,l}\frac{\delta V}{2} \left(e^{i(ql - \omega t)} + e^{-i(ql - \omega t)}\right) \hat{S}^z_{n,l}=  \sum\limits_{n,l}\delta V \mathrm{cos}(ql-\omega t) \hat{S}^z_{n,l}$. To get the first-order density response in the space-time domain, we add up the contributions
\begin{align}
   \rho_1 (n,l,t) &= \frac{1}{2} \rho_1(n,q, \omega) e^{i(ql- \omega t)}+\frac{1}{2} \rho_1(n,-q, -\omega) e^{-i(ql - \omega t)} \nonumber\\
    &= \frac{t_\parallel s_0^2\delta V\left(1-\mathrm{cos}(q)\right)}{\omega^2 -\omega^2_{1+}(q)} \mathrm{cos}(ql-\omega t),
\end{align}
and find the density modulation has a running wave profile. We can do the same for the longitudinal current
\begin{align}
    J_{1\parallel} (n,l,t) &= \frac{1}{2}J_{1\parallel}(n,q, \omega) e^{i(ql- \omega t)}+\frac{1}{2} J_{1\parallel}(n,-q, -\omega) e^{-i(ql - \omega t)} \nonumber\\
    &= \frac{t_\parallel s_0^2\omega \delta V}{2\left(\omega^2 -\omega^2_{1+}(q)\right)} \big(\mathrm{sin}(q) \mathrm{cos}(ql - \omega t) + (\mathrm{cos}(q) - 1)\mathrm{sin}(ql - \omega t)\big),
\end{align}
which is again a phase-shifted running wave, symmetric across the legs. The first-order density and current responses are invariant under a simultaneous sign change of $q$ and $\omega$. 

Alternatively, we could have a standing-wave drive: $\delta \hat{H} = \sum\limits_{n,l}\frac{\delta V}{4} \left(e^{i(ql - \omega t)} + e^{i(ql + \omega t)} + e^{i(-ql + \omega t)} + e^{i(-ql - \omega t)}\right) \hat{S}^z_{n,l}$, or rather $\delta \hat{H} = \sum\limits_{n,l}\delta V\mathrm{cos}(ql) \mathrm{cos}(\omega t)\hat{S}^z_{n,l}$. The induced first-order density modulation is then
\begin{align}
    \rho_1 (n,l,t)\rangle= \frac{t_\parallel s_0^2\delta V\left(1-\mathrm{cos}(q)\right)}{\omega^2 -\omega^2_{1+}(q)} \mathrm{cos}(ql) \mathrm{cos}(\omega t),
\end{align}
and the first-order current response is
\begin{align}
    J_{1\parallel} (n,l,t) =\frac{t_\parallel s_0^2\omega\delta V}{2\left( \omega^2 - \omega_{1+}^2(q)\right)}\big( \mathrm{sin}(q)\mathrm{sin}(ql) + (1-\mathrm{cos}(q))\mathrm{cos}(ql)\big)\mathrm{sin}(\omega t).
\end{align}
Both the density and current response are invariant under the sign change of either $q$ or $\omega$, as they should be for a standing wave. We see there is always a resonance when we drive at the dispersion of the symmetric collective mode.

One can proceed in a similar manner for second-order response. For a standing-wave drive, we saw that the first-order response has four peaks in the frequency-wavevector domain. Two-wave mixing will give all the possible peaks for the second-order response. Importantly, there will be four extremal peaks at $(\pm 2q,\pm 2\omega)$, which are the result of the wave $( q, \omega)$ mixing with $( q, \omega)$ or $( q, -\omega)$ with $( q, -\omega)$ and so forth. The four extremal peaks will all have the same magnitude by symmetry. The spatial profile of the second-order density response will include terms like $\mathrm{cos}(2ql)\mathrm{cos}(2\omega t)$. The prefactors of this terms will be a combination of the Fourier weights of the four extremal peaks. Therefore, it is sufficient to analyze the response to a monochromatic complex plane wave to predict the final magnitude of the second-order density modulations.

Lastly, we remark that while this appendix focuses on HCBs, the experimental considerations outlined here are generically applicable to lattice systems of bosons or fermions. The same holds true for the next section on Umklapp scattering.

\subsection{Umklapp scattering}
Since we are working with a lattice, the Brillouin zone extends from $-\pi$ to $\pi$ for momenta measured in units of the lattice length. If we drive at $|q|<\frac{\pi}{2}$, the nonlinear response $2q$ will still be inside the Brillouin zone. For drives at larger momenta, the wavevector will backfold to $2q-2\pi$, which is where the Fourier peak will be in experiment: the discrete Fourier transform of position space measurements cannot resolve signals oscillating faster than the lattice scale. For example, a drive at $q=\frac{3\pi}{4}$ induces second-order response at $2q\equiv -\frac{\pi}{2}$. Since a real experiment also drives at $-q$, there is signal at $\frac{\pi}{2}$ as well.

\subsection{Additional data}
\begin{figure}[h!]
        \centering
        \includegraphics[width = 0.5 \linewidth]{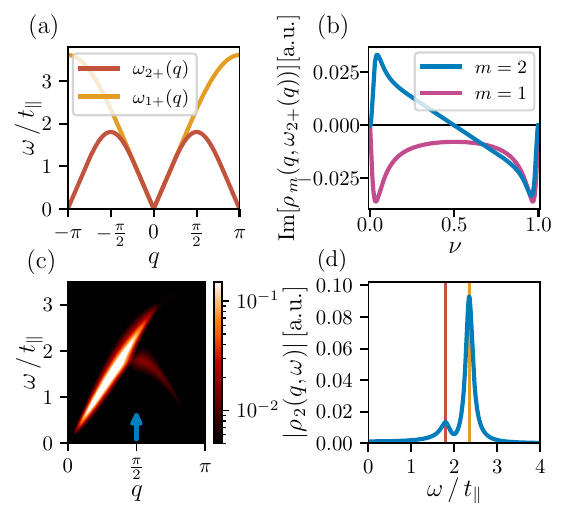}
        \caption{\textbf{Second harmonic generation on a HCB ladder.} (a) Dispersion of the linear and nonlinear collective modes $\omega_{1+}(q)$ and $\omega_{2+}(q)$. Data shown is for $t_\perp=t_\parallel$, damping rate $\eta=0.1t_\parallel$, driving strength $\delta V =0.2t_\parallel$ and filling $\nu=0.25$. (b) Imaginary part of the first- (magenta) and second-order density response at the nonlinear resonance $\omega =\omega_{2+}(\pi/2)$ for driving wavevector $q=\pi/2$, as a function of filling $\nu$. The strengths of the signals are comparable. At the half-filling point, the nonlinear signal flips sign, whereas the linear response is invariant under the particle-hole transformation. (c) Magnitude of the second-order density response $\vert \rho_2(q,\omega)\vert$ as a function of drive frequency and wavevector. The blue arrow denotes $q=\pi/2$, which is used in the next plot. (d) Nonlinear density response $\vert \rho_2(q,\omega)\vert$ at fixed wavevector $q=\pi/2$ is enhanced both when $\omega =\omega_{1+}(\pi/2)$ and when $\omega =\omega_{2+}(\pi/2)$. The y-label is shared between plots (c) and (d). }
        \label{fig:shg_app} 
\end{figure}

\begin{figure}[h!]
        \centering
        \includegraphics[width = 0.8 \linewidth]{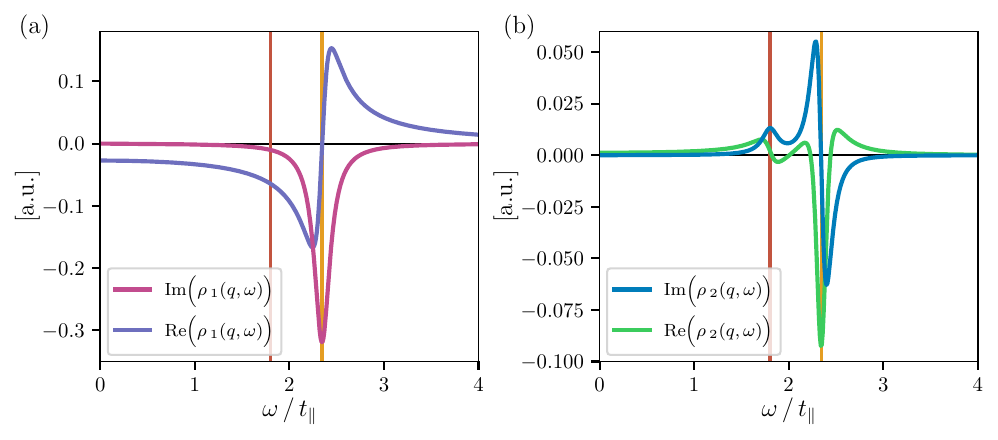}
        \caption{\textbf{Frequency dependence of the first- and second-order density modulations as a function of the drive frequency.} Data shown is for transverse hopping $t_\perp= t_\parallel$, damping rate $\eta=0.1t_\parallel$, driving strength $\delta V =0.2t_\parallel$ and filling $\nu=0.25$, at a fixed drive momentum $q=\pi/2$. Vertical lines denote the linear (orange) and nonlinear (brick red) resonance. (a) Imaginary and real part of the first-order density response. (b) Imaginary and real part of the second-order density response.}
        \label{fig:imre_app2} 
\end{figure}

In addition to the anisotropic ladder shown in the main text, we show results for a ladder of hard core bosons with $t_\perp = t_\parallel$ in Figures \ref{fig:shg_app} and \ref{fig:imre_app2}. We see the nonlinear peak at the linear resonance is higher while the linear peak is of similar height as for $t_\perp = 0.25t_\parallel$ data shown in the main text. 

\begin{figure}[h!]
        \centering
        \includegraphics[width = 0.7 \linewidth]{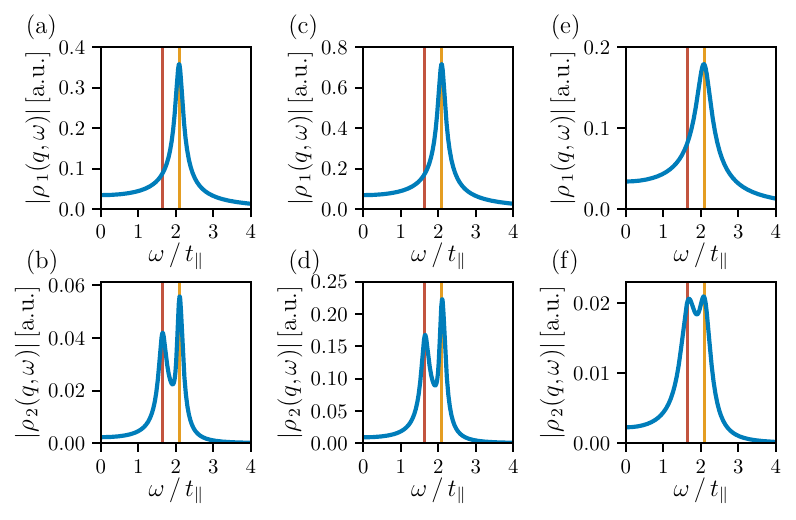}
        \caption{\textbf{Effects of driving strengths and damping rates on second harmonic generation.} All data shown is for a ladder of hard core bosons with $t_\perp=0.25 t_\parallel$ and filling $\nu=0.25$, at a fixed driving wavevector $q=\pi/2$. Vertical lines correspond to the linear (orange) and nonlinear (brick red) collective modes. (a) Linear density response $\vert \rho_1(q,\omega)\vert$ at driving strength $\delta V=0.2t_\parallel$ and damping rate $\eta=0.1 t_\parallel$. (b) Nonlinear density response $\vert \rho_2(q,\omega)\vert$ at $\delta V=0.2t_\parallel$ and $\eta=0.1 t_\parallel$. (c) Linear density response $\vert \rho_1(q,\omega)\vert$ at $\delta V=0.4 t_\parallel$ and $\eta=0.1 t_\parallel$ is twice as large as that in panel (a).  (d) Nonlinear density response $\vert \rho_2(q,\omega)\vert$ at $\delta V=0.4t_\parallel$ and $\eta=0.1 t_\parallel$ is four times as large as that in panel (b). (e) Linear density response $\vert \rho_1(q,\omega)\vert$ at $\delta V=0.2 t_\parallel$ and $\eta=0.2 t_\parallel$ is suppressed relative to that in panel (a).  (f) Nonlinear density response $\vert \rho_2(q,\omega)\vert$ at $\delta V=0.2t_\parallel$ and $\eta=0.2 t_\parallel$ is suppressed relative to that in panel (b).}
        \label{fig:shg_6pan_app} 
\end{figure}

We further investigate the effect of driving amplitude $\delta V$ and the dissipation rate $\eta$ in Figure \ref{fig:shg_6pan_app}. We see the linear signal doubles and the second-order signal quadruples when the driving amplitude is doubled, as expected from perturbation theory. Increasing the dissipation rate suppresses all peaks, both in first- and second-order response.

\newpage

\section{Finite temperature skewness}
In this section, we analyze in detail the density skewness protocol for fermionic systems. We first write out the grand-canonical skewness for a system of free fermions to gain insight about the relation between the Fermi-Dirac distribution function and density skewness.
We perform calculations for a grand-canonical system, with which we approximate a subsystem of the cold-atom lattice.
We then move to the Fermi-Hubbard model at finite temperature, which we treat with the mean-field spin-density wave (SDW) ansatz. We show that the skewness changes sign concomitantly with the onset of SDW order as either doping or temperature is varied.
This implies that general phase transitions that reorder the fermi surface may be accompanied by anomalies in the density skewness.

\subsection{Noninteracting fermion skewness}

First, we consider the case of free fermions with the Hamiltonian 
\begin{equation}
    \hat{H} = \sum_{a} [\epsilon_a -\mu]\hat{c}^\dagger_a \hat{c}_a ,
\end{equation}
where $a$ labels the energy eigenstates (with energies $\epsilon_a$) of the single-particle Hamiltonian, to be analyzed in the grand-canonical ensemble. 
We compute the cumulant-generating function for the total charge cumulants:
\begin{equation}
    C[\chi] = -i\frac{1}{V} \log \langle e^{i\chi \sum_a \hat{n}_a} \rangle. 
\end{equation}
We evaluate it as 
\begin{equation}
    C[\chi] = -i \frac{1}{V}\sum_\alpha\log \langle e^{i\chi \hat{n}_a}\rangle = -i\frac{1}{V}\sum_\alpha \log \left[ h_a + f_a e^{i\chi}\right],
\end{equation}
where 
\begin{equation}
    f_a =\frac{1}{ \exp(\beta[\epsilon_a - \mu])+1 }
\end{equation}
is the probability that eigenstate $a$ is occupied and $h_a = 1-f_a$ is the probability it is empty. 
Since each eigenmode can be taken as an independent Bernoulli random variable, the cumulant of the total number is the sum of the individual cumulants.
For a Bernoulli random variable, the skewness is $C^{(3)}_a = f_a h_a(h_a - f_a)$ and therefore we find the noninteracting result of 
\begin{equation}
    C^{(3)} = \frac{1}{V} \sum_a f_a (1-f_a) (1-2f_a).
\end{equation}
In the case of Bloch bands $\alpha$ with momentum $\mathbf{k}$, this reduces to 
\begin{equation}\label{app:c3free}
    C^{(3)} = \int_{\bf k} \sum_\alpha f_{\bf k\alpha} (1-f_{\bf k\alpha}) (1-2f_{\bf k \alpha}),
\end{equation}
where we have used the shorthand 
\begin{equation}
    \int_{\bf k} = \int_{-\pi}^{\pi} \frac{dk_x}{2\pi} \int_{-\pi}^{\pi} \frac{dk_y}{2\pi},
\end{equation}
valid for a two-dimensional square lattice.
The particle-hole asymmetry of the density skewness can be clearly seen from the $1-2f_{\bf k \alpha}$ in Eq. \eqref{app:c3free}.

\subsection{Mean-field theory}
In order to study the Fermi-Hubbard model at finite temperature and finite doping, we will use the spin-density wave (SDW) ansatz and treat the system within mean-field theory.
Though this is not well-justified, it is nevertheless an insightful and commonplace method~\cite{Schrieffer.1989,Chi.1994}, which can be made more formal by using dynamical mean-field theory (DMFT)~\cite{Peters.2014}. 
We start with the imaginary-time Matsubara action 
\begin{equation}
    \mathcal{S} = \int_0^\beta d\tau \left[\sum_{j\sigma} \overline{c}_{j\sigma}\left[ \partial_\tau - \mu + \frac12 U n_{j,\bar{\sigma}} \right]c_{j\sigma} - t \sum_{<j,k>\sigma} \overline{c}_{j\sigma} c_{k\sigma} + \textrm{h.c.}\right],
\end{equation}
which we will decouple by using a Gaussian ansatz for the Green's function.
First, we introduce the spin-density wave (SDW) spinors 
\begin{equation}
    \psi_{\bf p} = \begin{pmatrix}
        c_{\bf p\uparrow}(\tau) \\
        c_{\bf p + Q \downarrow }(\tau) 
    \end{pmatrix},
\end{equation}
where we have restricted ourselves for the moment to the commensurate order of SDW at $\mathbf{Q} = (\pi,\pi)$ with gap amplitude $\Delta$.
This is expected to adequately describe finite-range correlations in systems at finite temperature, at least when the magnetic order retains the commensurate order peak at $\mathbf{Q} =(\pi,\pi)$, which has been found to persist up to a finite threshold in doping~\cite{Yamada.1998}.
It may also be possible to generalize this ansatz to include incommensurate order as well. 

Now that we have introduced the SDW spinors, we can define the Matsubara Green's function
\begin{equation}
G(i\epsilon_m,\mathbf{p}) = -\frac{1}{\beta}\int \langle \psi_{\bf p}(\tau) \overline{\psi}_{\bf p}(0) \rangle e^{i\epsilon_m \tau}d\tau = \left[ i\epsilon_m +\mu - \epsilon_{\bf p} \tau_3 - \Delta \tau_1\right]^{-1}.
\end{equation}
Here $\hat{\tau}$ are the Pauli matrices in the SDW spinor space, $i\epsilon_m = 2\pi T (m+\frac12)$ are the fermionic Matsubara frequencies at temperature $T$, and $\epsilon_{\bf p} = -2t (\cos p_x + \cos p_y) $ is the square-lattice tight-binding dispersion.
Note that in principle, we can add a next-nearest-neighbor dispersion $t'$, leading to 
\begin{equation}
G(i\epsilon_m,\mathbf{p})  = \left[ i\epsilon_m +\mu -\epsilon'_{\bf k} - \epsilon_{\bf p} \tau_3 - \Delta \tau_1\right]^{-1},
\end{equation}
where 
\begin{equation}
\epsilon'_{\bf k} =-4 t' \cos k_x \cos k_y . 
\end{equation}
If we use the Feynman-Gibbs-Bogoliubov inequality for the free energy in terms of the Gaussian ansatz, variationally minimized over the SDW gap, we will recover the gap equation 
\begin{equation}
    \frac{1}{U} = -T\sum_{\epsilon_m}\int_{\bf p}\sum_\sigma \frac{1}{(i\epsilon_m +\mu)^2 - E_{\bf p\sigma}^2}
\end{equation}
for $\Delta$.
We also fix chemical potential in terms of doping via $n = 1-p = \int_{\bf p}\sum_{\sigma} f_{ \bf p \sigma}$, where $f_{\bf k \sigma} = [ e^{\beta E_{\bf k \sigma}}+1]^{-1}$ is the occupation function of the mode $\bf k\sigma$, $\sigma = \pm1$ and $E_{\bf k\sigma} = \sigma \sqrt{\epsilon_{\bf k}^2 + \Delta^2} -\mu$. 

\begin{figure}[t]
        \centering
        \includegraphics[width = 0.5 \linewidth]{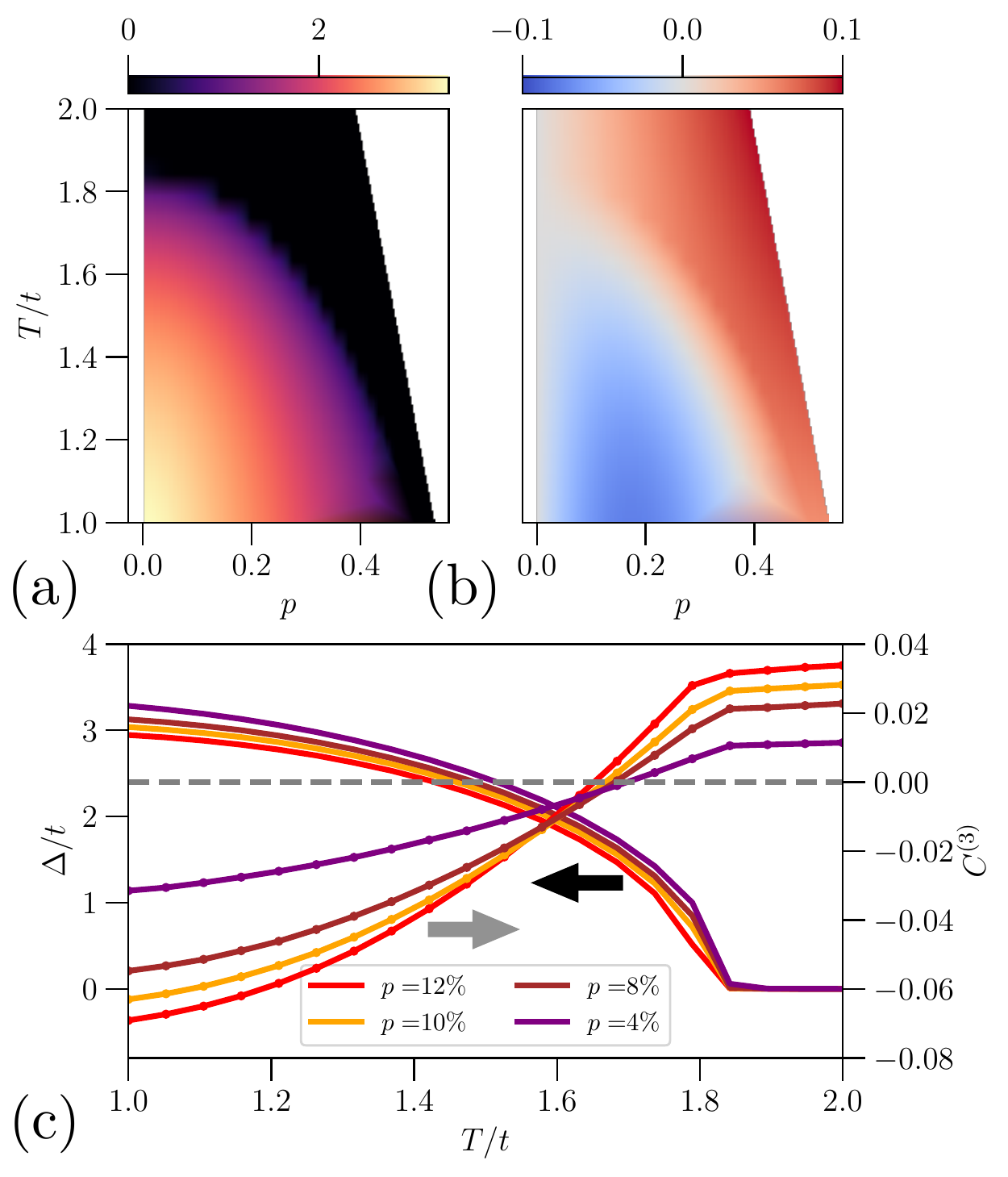}
        \caption{Skewness and SDW order at finite temperature. (a) Mean-field SDW order parameter for the Hubbard model with $U/t = 8$ at finite temperature and doping $p$. 
        We find a dome of SDW order near half-filling at low temperatures and near half-filling.
        (b) Skewness in the charge cumulants closely follows the shape of the SDW dome, which leads to a Fermi-surface reconstruction. 
        For dopings near half-filling, the skewness is hole-like rather than electron-like. 
        (c) At fixed small doping, we directly compare the onset of SDW order (left axis) with the change in density skewness (right axis), seeing that the density skewness closely follows the Neel order, but due to continuity onsets at a slightly lower temperature. 
        }
        \label{fig:skewness} 
\end{figure}

Due to the nesting instability on the square lattice, the Hubbard interaction leads to the onset of spin-density wave order with amplitude $\Delta$.
The SDW order is shown as a function of doping $p$ and temperature $T/t$ for $U/t = 8$ in Fig.~\ref{fig:skewness}(a), showing a clear dome of SDW order around half-filling $\delta = 1$ ($p=0$).
For $p = 0$ this extends to nearly $T_N/t \lesssim 2$, which is significantly higher than currently accessible temperatures in cold-atom simulators, which can currently reach $T/t \sim 0.1$. 
Finite doping then frustrates the nesting instability and ultimately leads to a melting of the commensurate magnetic order, though it is expected that incommensurate spiral orders will onset at finite doping~\cite{Peters.2014,Andrei.2026,Auerbach.1991,Shraiman.1989,Wietek.2021}, before ultimately melting at large hole-doping $p \to 1$.  
For simplicity, in this work we use a restricted commensurate SDW ansatz similar to Ref.~\cite{Schrieffer.1989}, though it would be interesting to explore the interplay of stripe or spiral orders with these higher cumulants. 

In order to compute the skewness, we use the free-fermion result from above. 
For non-interacting carriers we find this is positive when the doping is electron-like, and negative when the doping is hole-like. 
In Fig.~\ref{fig:skewness}(b) we see for $p>p_c$ the skewness is positive, as would be expected from a lightly-filled electron band.
On the other hand, within the dome of Neel order near half-filling ($p=0$), we see at low-temperatures the skewness is negative, indicating a low-temperature change to hole-like carriers.
This is consistent with the picture of the underdoped Mott insulator being better described by small hole-pockets being doped into an otherwise fully-filled lower Hubbard band, rather than a large electron-like Fermi surface. 
Within mean-field this is tied to the appearance of the SDW gap and therefore should be able to be seen in current quantum gas microscopes. 

This is more clearly seen in Fig.~\ref{fig:skewness}(c) where we fix the doping and study both the onset of SDW order and the behavior of the charge skewness. 
We can clearly see that the $C_3$ cumulant rapidly picks up on the Fermi surface reconstruction which starts at $T_N$.
Importantly, we note this analysis is in the absence of any explicit particle-hole symmetry breaking terms such as a next-nearest-neighbor hopping $t'$.
Including such a perturbation would be an interesting direction for future studies, as well as investigating the effect of incommensurate order. 

An intuitive picture of why the skewness is useful is that it is able to filter out contributions from the frozen charge which gives rise to local moments. 
Consider partitioning the total charge in the system into a frozen (bound) part $N_b$, which cannot fluctuate, and a free part $N_f$, which is itinerant and can fluctuate.
Then the average charge is simply $\langle N \rangle = N_b + \langle N_f \rangle $.
But the skewness of the charge is 
\begin{equation}
    C^{(3)} = \frac{1}{V} \langle N^3\rangle_c = \frac{1}{V} \langle \left( N - \langle N \rangle \right)^3 \rangle = \frac{1}{V} \langle \left( N_b + N_f - N_b - \langle N_f \rangle \right)^3 \rangle = \frac{1}{V}\langle (N_f - \langle N_f \rangle)^3 \rangle.
\end{equation}
Therefore, the skewness is insensitive to the contribution of frozen moments and only detects the particle-hole skewness of the itinerant carriers, which ultimately contribute to transport. 

Lastly, we speculate on the full temperature dependence of the subsystem skewness as a function of hole doping.
To obtain a qualitative picture for very small dopings in the magnetic polaron picture, we assume the subsystem skewness to resemble that of free particles, but with hole-like character rather than electron-like.
From \eqref{app:c3free} we see that for low temperatures, the magnitude of the skewness should increase with temperature. This would imply that at first, the skewness becomes more negative (hole-like) as temperature increases. 
However, the mean-field picture predicts that the trend should eventually reverse as the skewness crosses zero and becomes positive (electron-like) at high temperatures. The full temperature dependence for finite (and large) dopings and the crossover between different transport regimes is an exciting direction for future research.

\end{document}